\documentclass[twocolumn,trackchanges]{aastex631}

\usepackage{graphicx}
\usepackage{xcolor}
\usepackage{natbib}
\usepackage{amsmath}
\usepackage[subfigure]{tocloft}
\usepackage{subfigure}
\usepackage{booktabs}
\usepackage{CJK}
\usepackage{makecell}

\newcounter{RomanNumber}

\received{xxx}
\revised{xxx}
\accepted{xxx}
\shorttitle{Fate of the remnant in tidal stripping event}
\begin{document}
\begin{CJK*}{UTF8}{gbsn}

\title{Fate of the remnant in tidal stripping event: repeating and non-repeating}

\author[0000-0002-3525-791X]{Jin-Hong Chen (陈劲鸿)}
\affiliation{Department of Physics, University of Hong Kong, Pokfulam Road, Hong Kong, People's Republic of China}
\author[0000-0002-9589-5235]{Lixin Dai (戴丽心)}
\affiliation{Department of Physics, University of Hong Kong, Pokfulam Road, Hong Kong, People's Republic of China}
\author[0000-0002-9442-137X]{Shang-Fei Liu (刘尚飞)}
\affiliation{School of Physics and Astronomy, Sun Yat-sen University, Zhuhai 519082, China}
\affiliation{CSST Science Center for the Guangdong-Hong Kong-Macau Greater Bay Area, Sun Yat-sen University, Zhuhai 519082, China}
\author[0000-0002-6176-7745]{Jian-Wen Ou (欧建文)}
\affiliation{School of Intelligent Engineering, Shaoguan University, Shaoguan 512005, China}

\correspondingauthor{Jin-Hong Chen, Lixin Dai}
\email{chenjh2@hku.hk, lixindai@hku.hk}

%%%%%%%%%%%%%%%%%%%%%%%%%%%%%%%%%%%%%%%%%%%%%%%%%%%%%%%%%%%%
\begin{abstract}
Tidal disruption events (TDE) occur when a star ventures too close to a massive black hole. In a partial TDE (pTDE), the star only grazes the tidal radius, causing the outer envelope of the star to be stripped away while the stellar core survives. Previous research has shown that a star, once tidally stripped in a parabolic orbit, can acquire enough orbital energy for its remnant to become a high-velocity star potentially capable of escaping the galaxy. Conversely, some studies have reported that the remnant may lose orbital energy and undergo re-disruption, leading to a recurring pTDE. This study aims to uncover the physical mechanisms and determine the conditions that lead to these divergent outcomes. We find that the orbital energy change only depends on the impact factor and the stellar structure, and barely depends on the mass of the black hole or the exact mass or orbital eccentricity of the star. For a $\gamma=5/3$ (or $\gamma=4/3$) polytropic star, after a pTDE its remnant gains orbital energy when the impact factor $\beta \gtrsim 0.62$ (or $\gtrsim 1.1$) or loses energy vice versa. Additionally, we verify an analytical equation for orbital energy change that is applicable across various systems. Through hydrodynamic simulations, we also explore the structure of the stellar remnant post-tidal stripping. Our findings provide critical insights for interpreting observed pTDEs and advancing knowledge on the orbital evolution and event rate of these events.
\end{abstract}


%%%%%%%%%%%%%%%%%%%%%%%%%%%%%%%%%%%%%%%%%%%%%%%%%%%%%%%%
\section{Introduction}
\label{sec:introduction}
A star can be occasionally disturbed by nearby stars, causing it enter an orbit where its pericenter is too close to a central massive black hole (MBH). This proximity can lead to the star being disrupted by the tidal force. The subsequent accretion of material falling back onto the MBH produces a bright flare that illuminates the galaxy \citep{Rees_Tidal_1988}. These events, known as tidal disruption events (TDEs), can be used to discover and study supermassive black holes (SMBHs, $10^{6 - 8}\ M_{\odot}$) and intermediate-mass black holes (IMBH, $10^{2 - 5}\ M_{\odot}$), which are typically found in the centers of galaxies and star clusters \citep{Miller_Production_2002,Portegies_The_2002,Greene_IMBH_2020}, respectively. 

If the tidal force is not strong enough to completely disrupt the star, a remnant may survive, with only a small fraction of its mass stripped away during the encounter. These partial tidal disruption events (pTDEs) can occur more frequently than full TDEs \citep{Chen_pTDE_2021,Bortolas_pTDE_2023}. Some TDE candidates, e.g., AT 2023clx \citep{zhong_black_2024}, AT 2018hyz \citep{Gomez_AT2018hyz_2020}, AT 2019qiz \citep{Nicholl_AT2019qiz_2020} and iPTD16fnl \citep{Blagorodnova_iPTF16fnl_2017}, are thought to be the possible pTDEs based on their relatively low luminosity or fast-evolving light curves.

If the orbit of the surviving remnant remains bound to the MBH, it will return to the pericenter and be tidally disrupted again. Several repeated pTDE candidates have been discovered through optical and X-ray surveys, e.g., AT 2020vdq \citep{Somalwar_first_2023}, ASASSN-14ko \citep{Payne_14ko_2021,Payne_14ko_2022,Payne_chandra_2023}, AT 2018fyk \citep{Wevers_2018fyk_2023}, eRASSt J045650.3-203750 \citep{Liu_J045650_2023}, AT 2022dbl \citep{lin_unluckiest_2024}, RX J133157.6-324319.7 \citep{Hampel_new_2022,Malyali_rebrightening_2023} and IRAS F01004-2237 \citep{sun_recurring_2024}. There are also other similar events, such as quasi-periodic eruptions (QPEs) in galactic nuclei \citep{Miniutti_Nine_2019,Giustini_Xray_2020,Arcodia_QPE_2021,Chakraborty_Possible_2021} and ultraluminous X-ray bursts in star clusters \citep{Sivakoff_Luminous_2005,Jonker_Discovery_2013,Irwin_Ultraluminous_2016}, which are believed to originate from the repeated tidal stripping of a white dwarf by an IMBH \citep{Shen_Fast_2019,Zhao_QPE_2021,Wang_a_2022,king_QPE_2022,Chen_tidal_2023,Lu_QPE_2023}.

One particularly interesting case is ASASSN-14ko, which exhibits firm periodic behavior ($P \simeq 115$ days) with a periodic derivative of $\dot P \simeq -0.0026$, suggesting orbital shrinkage \citep{Payne_chandra_2023}. It has been suggested that the orbital shrinkage of ASASSN-14ko could be caused by the tidal stripping process \citep{Ryu_tidal3_2020,Payne_14ko_2021}, although further exploration is needed. Tidal interaction and the mass loss during the encounter near the pericenter can alter the orbital energy and stellar structure of the remnant. This effect is not only crucial for the orbital evolution of pTDEs, but also can contributes to the observational differences between each flare.

The orbital change of the stellar remnant following pTDEs has also been considered as a mechanism for ejecting stars and turning them into high-velocity stars that may escape the host galaxy \citep{manukian_turbovelocity_2013,Gafton_relativistic_2015}. However, \cite{Ryu_tidal3_2020} found that in some cases of their simulations, the remnants can lose orbital energy and become bound to SMBHs instead of gaining energy. Similarly, \cite{Kiroglu_partial_2023} study the pTDEs caused by IMBHs and also found cases where the remnants lose energy. Despite these findings, a robust interpretation of these phenomena remains elusive, likely due to variations in simulation codes, stellar structures, and other parameters across studies. To address this, we aim to conduct a systematic investigation to determine the exact conditions and parameter ranges that dictate whether the remnant's orbit will gain or lose energy.

With the launch of many time-domain telescopes and the advancement of sensitive all-sky surveys and monitoring in optical, ultraviolet and X-ray bands, studying TDEs, especially the repeated ones, is currently in a golden era. The observational properties of these systems depend on orbital parameters and the stellar internal structure, making dedicated investigations into the orbital and structural changes of the remnants is necessary. 

To study pTDEs, we simulate the partial disruption processes involving MBH (both SMBH and IMBH) using a 3D hydrodynamic simulation code (FLASH) (section \ref{subsec:simulation_setup}). We present and analyse the results concerning orbital changes and stellar structure alterations in section \ref{subsec:simulation_result}. Additionally, we explore the underlying physical mechanisms behind the orbital energy changes in \ref{sec:bound_unbound}, identifying the parameter range for energy loss and gain. Base on our simulation results, we predict the subsequent passages of repeating pTDEs in Section \ref{sec:pTDE}. We discuss the pTDE candidates and its rate in Section \ref{sec:discussion}, and conclude with a summary of our findings in Section \ref{sec:conclusion}.

%%%%%%%%%%%%%%%%%%%%%%%%%%%%%%%%%%%%%%%%%%%%%%%%%%%%%

%%%%%%%%%%%%%%%%%%%%%%%%%%%%%%%%%%%%%%%%%
\section{Method}
\label{sec:method}
\subsection{3D hydrodynamic simulation setup}
\label{subsec:simulation_setup}
We conduct simulations of pTDEs using FLASH (version 4.0), an adaptive-mesh, grid-based hydrodynamics code \citep{Fryxell_FLASH_2000}. The hydrodynamic implementation we used is the directionally split piecewise-parabolic approach \citep{Colella_PPM_1984}, provided within the FLASH framework. We adopt the modified gravity algorithm from \cite{Guillochon_Consequences_2011} and used the same multipole gravity solver settings as \cite{Guillochon_Hydrodynamical_2013}. The simulations are preformed in the rest frame of the star.

The stars are modeled as polytropes, with the polytropic index $\gamma$ set to either $5/3$ or $4/3$. These 1D profiles are then mapped to the 3D Cartesian grid. The star is relaxed for at least five times its dynamical timescale $t_{\rm dyn} \simeq \sqrt{R_*^3/(GM_*)}$, where $M_*$ and $R_*$ are the stellar mass and radius, respectively. The simulation domain is a cubic volume with a size of $4 \times 10^{14}$ cm, initially composed of a single $8^3$ block, which is then bisected into smaller $8$ blocks up to 15 times based on the PARAMESH refinement criteria \citep{lohner_adaptive_1987}, resulting in a minimum cell size of $ \sim 0.02\ R_{\odot}$ \footnote{To get a precise orbital change, we enhance the resolution of the simulation with $\gamma=4/3$, $\beta=0.6$ to $0.005\ R_{\odot}$.}. 

At the beginning of each simulation, the star is placed at a distance of five times the tidal radius from the central MBH, i.e., $5 \times R_{\rm t} = 5R_* (M_{\rm h}/M_*)^{1/3}$, and allowed to move inward along a parabolic or eccentric orbit, where $M_{\rm h}$ is the MBH mass. During the simulation, the gas evolves adiabatically according to the equation of state $\Gamma = d\ln{P}/d\ln{\rho} = 5/3$. As the star passes near the pericenter, the gravitational force of the MBH tidally strips the star, elongating the debris stream on both sides. For the simulations with the star on a parabolic orbit, we end the simulation when the star moves to $50R_{\rm t}$ away from the MBH; for the eccentric case, we end it when the star is close to the apocenter.

The goal of this paper is to study the orbital change of the remnant and its dependence on various parameters. Thus, we conduct dozens simulations with different orbital parameters and MBH masses, ranging from scenarios of minimal mass loss to full disruption.

\begin{table}
\centering
\caption{Simulation results of stripped mass and orbital change across various parameters.}
\label{tab:sim_results}
\begin{tabular}{ccccccc}
\toprule
$\gamma$ & $M_{\rm h}$ & $M_*$ & $\beta$ & $e$ & $\Delta M$ & $\Delta \epsilon_{\rm orb}$ \\
 & $(M_{\odot})$ & $(M_{\odot})$ & & & $(M_{\odot})$ & (erg/g) \\
\midrule
$4/3$ & $10^6$ & 1 & $0.4$ & $1$ & N/A & N/A \\
$4/3$ & $10^6$ & 1 & $0.5$ & $1$ & N/A & N/A \\
$4/3$ & $10^6$ & 1 & $0.6$ & $1$ & $8.2 \times 10^{-5}$ & $-3.1 \times 10^{12}$ \\
$4/3$ & $10^6$ & 1 & $0.7$ & $1$ & $2.3 \times 10^{-3}$ & $-3.7 \times 10^{12}$ \\
$4/3$ & $10^6$ & 1 & $0.8$ & $1$ & $10^{-2}$ & $-8.1 \times 10^{12}$ \\
$4/3$ & $10^6$ & 1 & $0.85$ & $1$ & $2 \times 10^{-2}$ & $-2 \times 10^{13}$ \\
$4/3$ & $10^6$ & 1 & $0.9$ & $1$ & $3.4 \times 10^{-2}$ & $-2.2 \times 10^{13}$ \\
$4/3$ & $10^6$ & 1 & $1.0$ & $1$ & $7.2 \times 10^{-2}$ & $-3.1 \times 10^{13}$ \\
$4/3$ & $10^6$ & 1 & $1.1$ & $1$ & $0.13$ & $-8.7 \times 10^{12}$ \\
$4/3$ & $10^6$ & 1 & $1.2$ & $1$ & $0.21$ & $2.8 \times 10^{13}$ \\
$4/3$ & $10^6$ & 1 & $1.3$ & $1$ & $0.30$ & $8.9 \times 10^{13}$ \\
$4/3$ & $10^6$ & 1 & $1.4$ & $1$ & $0.40$ & $1.6 \times 10^{14}$ \\
$4/3$ & $10^6$ & 1 & $1.5$ & $1$ & $0.50$ & $3.1 \times 10^{14}$ \\
$4/3$ & $10^6$ & 1 & $1.55$ & $1$ & $0.55$ & $3.8 \times 10^{14}$ \\
$4/3$ & $10^6$ & 1 & $1.6$ & $1$ & $0.61$ & $5.7 \times 10^{14}$ \\
\midrule
$4/3$ & \textcolor{red}{$10^3$} & 1 & $1.0$ & $1$ & $7.3\times10^{-2}$ & $-1.3 \times 10^{13}$ \\
$4/3$ & \textcolor{red}{$10^3$} & 1 & $1.3$ & $1$ & $0.30$ & $1.0 \times 10^{14}$ \\
$4/3$ & $10^6$ & 1 & $1.3$ & \textcolor{red}{$0.9$} & $0.39$ & $3.5 \times 10^{14}$ \\
\midrule
$5/3$ & $10^6$ & 1 & $0.4$ & $1$ & N/A & N/A \\
$5/3$ & $10^6$ & 1 & $0.5$ & $1$ & $1.7 \times 10^{-4}$ & $-3.5 \times 10^{13}$ \\
$5/3$ & $10^6$ & 1 & $0.55$ & $1$ & $8.7 \times 10^{-3}$ & $-5 \times 10^{13}$ \\
$5/3$ & $10^6$ & 1 & $0.6$ & $1$ & $4.7 \times 10^{-2}$ & $-2.6 \times 10^{13}$ \\
$5/3$ & $10^6$ & 1 & $0.65$ & $1$ & $0.13$ & $5.2 \times 10^{13}$  \\
$5/3$ & $10^6$ & 1 & $0.7$ & $1$ & $0.24$ & $2 \times 10^{14}$  \\
$5/3$ & $10^6$ & 1 & $0.75$ & $1$ & $0.38$ & $3.8 \times 10^{14}$  \\
$5/3$ & $10^6$ & 1 & $0.77$ & $1$ & $0.44$ & $4.8 \times 10^{14}$  \\
$5/3$ & $10^6$ & 1 & $0.8$ & $1$ & $0.54$ & $6.4 \times 10^{14}$  \\
\midrule
$5/3$ & \textcolor{red}{$10^3$} & 1 & $0.55$ & $1$ & $9.8 \times 10^{-3}$ & $-4.7 \times 10^{13}$ \\
$5/3$ & \textcolor{red}{$10^3$} & 1 & $0.7$ & $1$ & $0.24$ & $1.7 \times 10^{14}$ \\
$5/3$ & $10^6$ & \textcolor{red}{0.1} & $0.55$ & $1$ & $8.7 \times 10^{-3}$ & $-3.6 \times 10^{13}$ \\
$5/3$ & $10^6$ & \textcolor{red}{0.1} & $0.7$ & $1$ & $0.24$ & $7.3 \times 10^{13}$ \\
$5/3$ & $10^6$ & 1 & $0.55$ & \textcolor{red}{$0.9$} & $1.4 \times 10^{-2}$ & $-3.6 \times 10^{13}$ \\
\bottomrule
\end{tabular}
\tablecomments{\footnotesize In columns 1--5, we list the initial parameters for each simulation: stellar polytrope, MBH mass, stellar mass, impact factor, and eccentricity, respectively. The final results, including the amount of mass loss and the change of specific orbital energy of the remnant, are listed in columns 6--7. For small $\beta$ where $\Delta M \lesssim 10^{-6}\ M_{\odot}$, no significant mass loss is observed, thus we mark these cases as N/A. Our resolution is insufficient to determine $\Delta \epsilon_{\rm orb}$ for small $\beta \lesssim 0.5$, thus we also mark these as N/A. The default setup for most of our simulations is $M_{\rm h} = 10^6\ M_{\odot}$, $M_* = M_{\odot}$, $R_*= R_{\odot}$ and $e=1$. Parameters with different setups for studying parameter dependence are highlighted in red. For simulations with $M_* = 0.1\ M_{\odot}$, the corresponding stellar radius is $R_* = 0.16\ R_{\odot}$ \citep{Kippenhahn_Stellar_1994}.}
\end{table}

\begin{figure}		
\centering
    \includegraphics[scale=0.17]{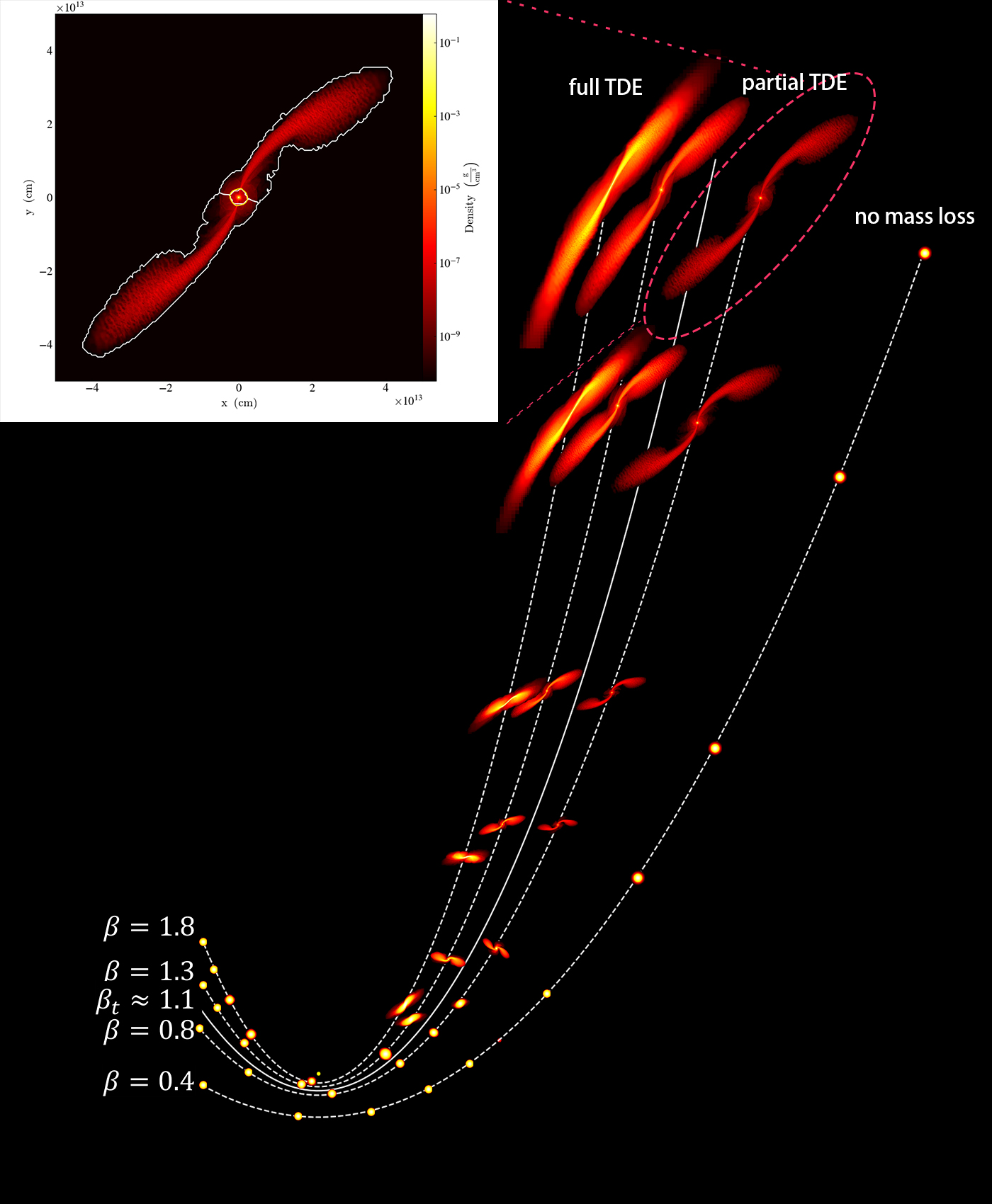}
    \caption{Snapshots from multiple simulations depicting the tidal disruption of a polytropic star ($\gamma = 4/3$, $M_* = M_{\odot}$, $R_* = M_{\odot}$, $M_{\rm h}=10^6\ M_{\odot}$, $e=1$) and with different $\beta$ parameters. The sizes of the snapshots have been adjusted for clarity. In cases with large pericenters, e.g., $\beta = 0.4$, the tidal force is too weak to cause a TDE, but it can induce stellar oscillation. For intermediate $\beta$, the tidal force strips away the outer layers of the star after the encounter, producing a pTDE. For large $\beta$, the tidal force is strong enough to fully destroy the star. The orbits with different $\beta$ are indicated by the white lines, and that with the transitional impact factor $\beta_{\rm t} \sim 1.1$ is in between. When $\beta \lesssim \beta_{\rm t}$, the dominance of tidal excitation over orbital change results in the remnant losing orbital energy after the encounter. Conversely, if $\beta \gtrsim \beta_{\rm t}$, the dominance of asymmetric mass loss leads to the remnant gaining orbital energy after the encounter. The inset panel shows the projection of the gas density on the orbital plane with $\beta = 0.8$ at $t=3 \times 10^5~$s since the beginning of the simulation. The material enclosed in the yellow contour is the remnant. The left and right parts of the stream, enclosed by the white lines, are the stripped mass, which are bound and unbound to the central MBH, respectively.}
    \label{fig:orbit_sketch}
\end{figure}

We run simulations for different impact parameters $\beta \equiv R_{\rm t}/R_{\rm p}$ ranging from $\beta = 0.4$ to $1.6$ for $\gamma = 4/3$, and 0.4 to 0.8 for $\gamma = 5/3$, where $R_{\rm p}$ is the stellar orbital pericenter radius. For large pericenter radius (small $\beta$), the tidal force is too weak to cause a TDE. Conversely, for large $\beta$, the entire star can be fully disrupted by the MBH. Snapshots of several simulations are shown in Figure \ref{fig:orbit_sketch} to illustrate different scenarios. 

The stellar structure with $\gamma = 4/3$ is more centrally concentrated than that of $\gamma =5/3$. Therefore, a star with $\gamma = 4/3$ is more difficult to full disrupt. The critical $\beta_c$ for full TDEs is $0.9$ for $\gamma = 5/3$ and $1.85$ for $\gamma =4/3$ \citep{Guillochon_Hydrodynamical_2013}.

In most of the simulations, we use a standard polytropic star with solar mass and radius. Given that most tidally disrupted stars are likely from the lower end of the stellar mass distribution, we also simulate the tidal disruption of a low mass star with $M_* = 0.1\ M_{\odot}$ and $R_* = 0.16\ R_{\odot}$, corresponding to the mass-radius relation of a main-sequence star \citep{Kippenhahn_Stellar_1994}. The polytropic index for the low mass star is set to be $\gamma = 5/3$, approximating to the profile of a convection-dominated star \citep{golightly_diversity_2019}.

We focus on TDEs involving both SMBHs and IMBHs, setting the MBH mass to $10^6\ M_{\odot}$ and $10^3\ M_{\odot}$ in the simulations. The stars are initially on parabolic orbits ($e=1$) and eccentric orbits ($e =0.9$), corresponding to one-off TDEs and repeating TDEs, respectively. The parameters and results are listed in Table \ref{tab:sim_results}. The default setup of the simulation is $M_{\rm h} = 10^6\ M_{\odot}$, $M_* = M_{\odot}$ and $e=1$.

\subsection{Simulation results}
\label{subsec:simulation_result}

\subsubsection{Tidally stripped mass}
\label{subsubsec:stripped_mass}
To estimate which gas belongs to the remnant, we adopt an iterative energy-based approach as outlined in \cite {Guillochon_Hydrodynamical_2013}. First, we calculate the specific binding energy of material in a given cell using
\begin{equation} \label{eq:binding_energy}
    E_{{\rm b}, i} = 0.5 (\vec{v}_i - \vec{v}_0)^2 + \phi_i,
\end{equation}
where $v_i$ and $\phi_i$ are the gas velocity and gravitational self-potential in a given cell, respectively, with $\phi_i$ calculated by the multipole solver. The reference velocity $v_0$ is initially taken to be the velocity at the location of the star’s peak density. Next, we sum over all mass elements for which have $E_{{\rm b}, i} < 0$. Third, we re-evaluate the reference velocity $v_0$ to be the new center of momentum. This process is repeated until $v_0$ converges to a constant value. 

\begin{figure}		
\centering
 \includegraphics[scale=0.5]{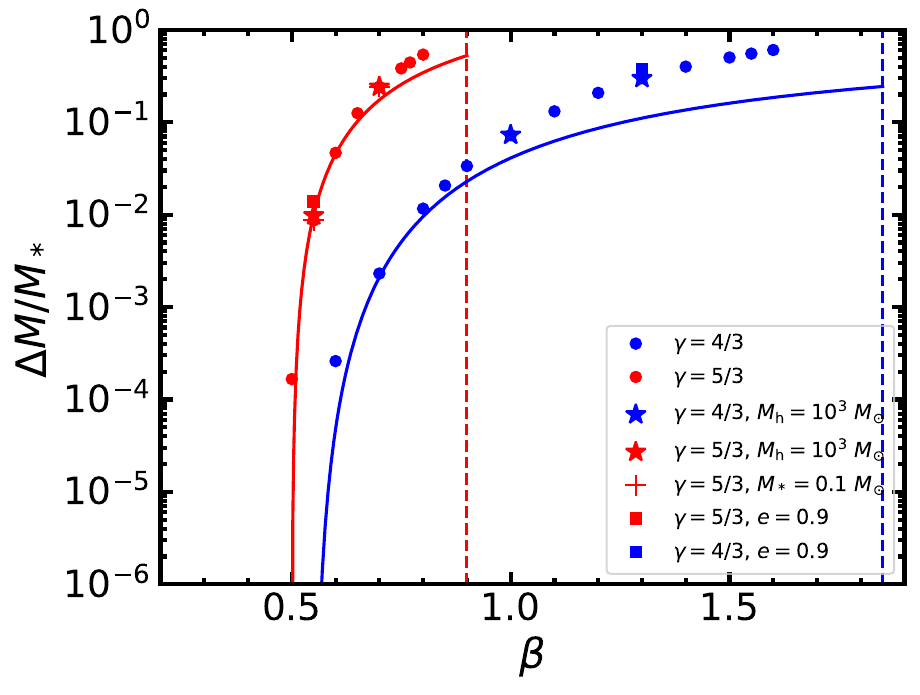}
\caption{Tidally stripped stellar mass versus $\beta$. The data points are the results at the end of simulations. The solid lines depict the analytical estimate given by Eq. (\ref{eq:dm_beta}). The vertical dashed lines indicate the critical  $\beta$ for full TDEs, which are $\beta_c=0.9$ for $\gamma = 5/3$ and and $\beta_c=1.85$ for $\gamma = 4/3$ \citep{Guillochon_Hydrodynamical_2013}. The circles, stars, crosses, and square points represent the simulations with default setup ($M_{\rm h}=10^6\ M_{\odot}$, $M_* = M_{\odot}$,$R_* = R_{\odot}$, $e=1$), with IMBH ($M_{\rm h} = 10^3\ M_{\odot}$), with a small star ($M_* = 0.1\ M_{\odot}$, $R_* = 0.16\ R_{\odot}$), and on eccentric orbit ($e=0.9$), respectively. Overall, $\Delta M/M_*$ depends only on $\beta$ and the stellar structure $\gamma$, but not on the MBH mass, stellar mass, or stellar orbital eccentricity, consistent with the analytical result.}
\label{fig:dm_beta}
\end{figure}

When $\beta \lesssim 0.5$, the tidally stripped mass is $\lesssim 10^{-6}\ M_{\odot}$ for both of $\gamma = 4/3$ and $\gamma = 5/3$, indicating negligible mass loss in these cases. The stripped mass $\Delta M$ versus $\beta$ is shown in Figure \ref{fig:dm_beta}, where $\Delta M$ increases steeply with larger $\beta$, as expected.

To provide a physical interpretation of this trend, we analytically estimate the relation between $\Delta M$ and $\beta$. At the pericenter, the star momentarily overflows its Roche lobe $R_{\rm lobe} \simeq \beta_0 R_{\rm p}(M_{\rm h}/M_*)^{-1/3}$, which strips away an exterior layer. The exact value of $\beta_0$ depends on the stellar structure, orbital eccentricity, and the mass ratio $M_{\rm h}/M_*$ \citep{Sepinsky_Equipotential_2007}. We take $\beta_0 = 0.5$ and $0.55$ for $\gamma = 5/3$ and $\gamma = 4/3$, respectively, according to our simulation outcomes.

Assuming the stripped layer is a spherical shell at the stellar surface with depth $z(\beta) = R_* - R_{\rm lobe} \simeq (1-\beta_0/\beta)R_*$, we estimate the stripped mass as 
\begin{equation} \label{eq:dm1}
    \Delta M(z) \simeq 4\pi R_*^2 \int^z_0 \rho (z') dz',
\end{equation}
where $\rho$ is the gas density. Using the hydrostatic equation at the surface $dP/dr \simeq -G\rho M_*/R_*^2$, where $P$ is the pressure, and substituting the polytropic relation $P = \mathrm{K} \rho^{\gamma}$, with $\mathrm{K}$ given by \citep{chandrasekhar_introduction_1939}
\begin{equation} \label{eq:kappa}
    \mathrm{K} = 4\pi G\frac{\gamma-1}{\gamma} \rho_{\rm c}^{2-\gamma} \left(\frac{R_*}{\xi_{\rm n}} \right)^2,
\end{equation}
we obtain 
\begin{equation} \label{eq:dm_beta}
\begin{split}
    \frac{\Delta M}{M_*} &\simeq \frac{\gamma-1}{\gamma} \xi_n^{2/(\gamma-1)} \left(3\frac{\rho_{\rm c}}{\overline{\rho}_*} \right)^{\frac{\gamma-2}{\gamma-1}} \left(1-\frac{\beta_0}{\beta} \right)^{\gamma/(\gamma-1)} \\
    &\simeq \begin{cases}
        4\left(1-\frac{0.5}{\beta} \right)^{5/2},\quad \gamma=5/3 \\
        \left(1-\frac{0.55}{\beta} \right)^{4},\quad \gamma=4/3.
    \end{cases}
\end{split}
\end{equation}
For $\gamma = 4/3$ and $5/3$ polytropic stars, the dimensionless radius is $\xi_{\rm n} = 6.9$ and $3.6$, and the ratio of central density to average density is $\rho_c/\overline{\rho}_* = 54.2$ and $6$, respectively.

Figure \ref{fig:dm_beta} shows that the analytical estimate closely matches the simulation results. However, for large $\beta$, when the tidal force is strong and tidal deformation is significant, Eq. (\ref{eq:dm_beta}) underestimates the stripped mass.

Even though our simulation results for the stripped mass are consistent with \cite{Guillochon_Hydrodynamical_2013} and analytical estimates, the stripped mass for small $\beta \sim 0.5$ shows larger discrepancies between results obtained from different simulation codes, as noted in Figures 4 and 5 of \cite{Mainetti_The_2017}. This discrepancy may be due to differences in resolution or the methods employed by the various codes. To obtain an accurate value for the stripped mass, the cell size should be smaller than or at least comparable to the depth of the stripped layer. Using Eq. (\ref{eq:dm1}), we estimate the depth of the stripped layer to be $z \sim 0.6(\Delta M/M_*)^{2/5}\ R_*$ and $z \sim (\Delta M/M_*)^{1/4}\ R_*$ for $\gamma = 5/3$ and $4/3$, respectively, which gives $z \sim 0.02\ R_*$ and $0.1\ R_*$ for $\Delta M/M_* \simeq 10^{-4}$. Thus, our simulation resolution is at the limit for $\gamma = 5/3$ and more than sufficient for $\gamma = 4/3$. Weak pTDEs with small $\beta$ require further investigation with higher resolution simulations.

\subsubsection{Remnant's orbit}
\label{subsubsec:orbit}
As we determine the location and velocity of the stellar remnant, we can further estimate the remnant's orbital motion, such as the relative velocity $\vec{v}_{\rm rel}$ and separation distance $\vec{R}_{\rm rel}$ between the remnant and the MBH. We calculate the orbital energy of the remnant using
\begin{equation} \label{eq:E_orb}
    E_{\rm orb} = \frac12 \mu \vec{v}_{\rm rel}^2 - \frac{G (M_{\rm h}+M_{\rm rem}) \mu}{|\vec{R}_{\rm rel}|},
\end{equation}
where $M_{\rm rem} = M_* - \Delta M$ and $\mu = M_{\rm rem} M_{\rm h}/(M_{\rm h}+M_{\rm rem})$ are remnant mass and the reduced mass, respectively. The specific energy is $\epsilon_{\rm orb} = E_{\rm orb}/\mu$.

The specific orbital angular momentum of the remnant is calculated by 
\begin{equation} \label{eq:j_orb}
    \vec{j}_{\rm orb} = \vec{R}_{\rm rel} \times \vec{v}_{\rm rel}.
\end{equation}

During the tidal stripping process, the orbital energy and the angular momentum of the remnant change continuously due to two processes: 1) tidal excitation; 2) mass loss due to tidal stripping. 

\begin{figure*}	
\centering
 \includegraphics[scale=0.4]{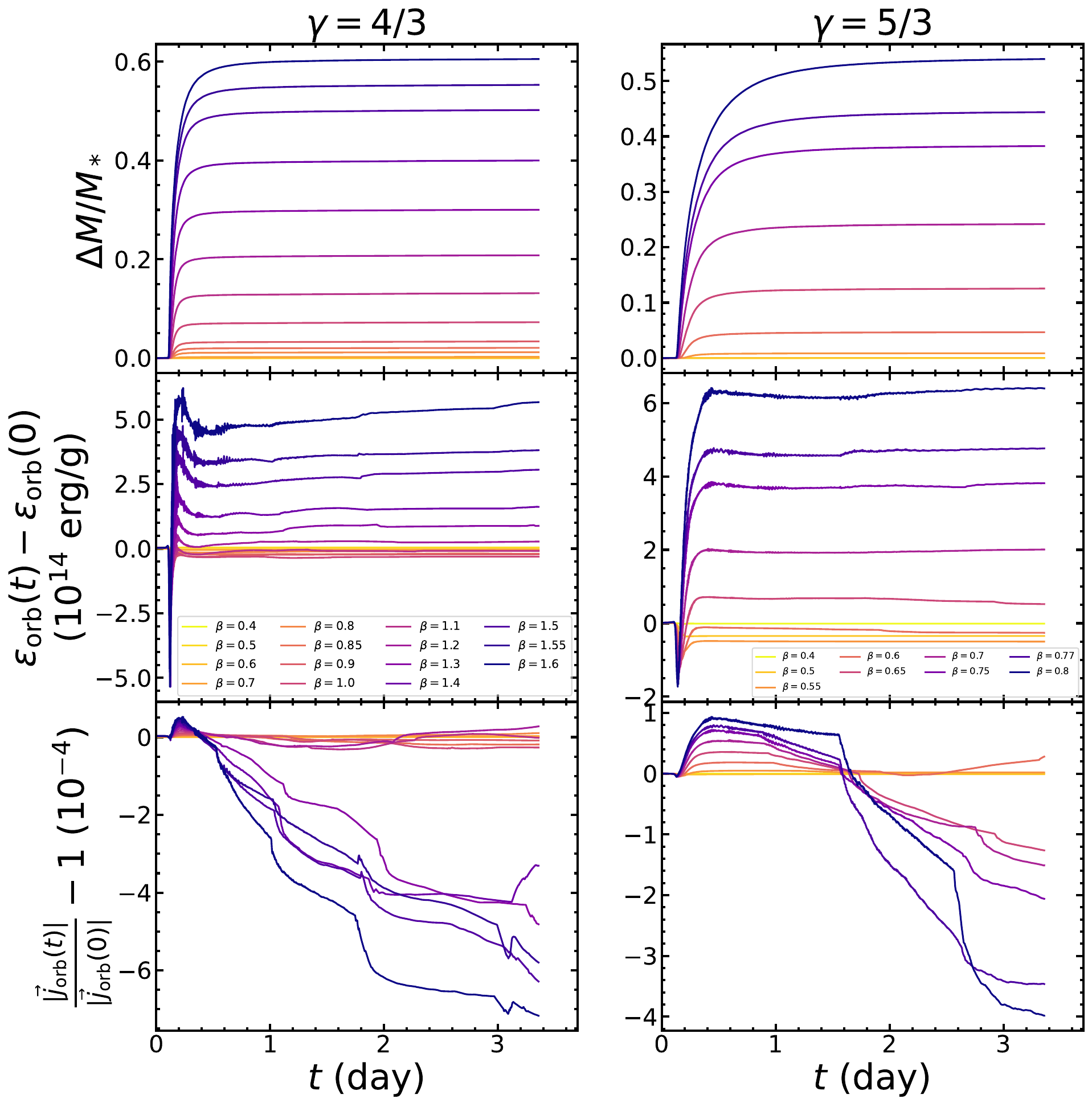}
\caption{Evolution of stripped stellar mass, remnant specific orbital energy, and remnant specific angular momentum are shown in upper, middle, and lower panels, respectively. The left and right figures depict results with $\gamma = 4/3$ and $\gamma = 5/3$ polytropic stars, respectively, using the default setup ($M_{\rm h}=10^6\ M_{\odot}$, $M_* = M_{\odot}$, $R_*= R_{\odot}$, $e=1$). Curves of different colors correspond to different $\beta$ values. The specific energy suddenly drops when the star passes by the pericenter radius, then rises and attains an almost constant value at late time. This occurs because part of the orbital energy is dissipated and transferred to the oscillations of the remnant. At late time, asymmetric mass loss causes the remnant to gain some energy; we will discuss the details in section \ref{subsec:osc_strip}. We find that the remnant's mass and its orbit has attained a stable value at the end of the simulations. The specific angular momentum of the remnant has negligible change, especially for those cases where the remnants are bound to the MBHs.}
\label{fig:Eorb_Rp_dM}
\end{figure*}

Figure \ref{fig:Eorb_Rp_dM} shows the evolution of these quantities along with the fractional stripped mass. After the star passes the pericenter ($t\sim 0.2$ day), it starts to be rapidly stripped by the tidal force. As the remnant moves away from the MBH, the stripped mass gradually stabilizes, and its orbit becomes stable. Our simulations indicate that both the stripped mass and the remnant's orbit achieve stable values by the end of the simulations. While the orbital angular momentum $\vec{j}_{\rm orb}$ for cases with large $\beta$ still appears to evolve slightly at the end of the simulation, the overall change is minimal ($\lesssim 10^{-4}$) and can be considered negligible.

Initially, the orbital energy of the remnant decreases and then increases to a constant value. This behavior occurs because the tidal force overcomes the self-gravity, inducing tidal oscillations that dissipate some of the orbital energy. Later, as the stripped mass flows away from the remnant, the redistribution of energy causes the remnant to gain some orbital energy. These effects are discussed in detail in section \ref{subsec:osc_strip}.

%Using $\epsilon_{\rm orb}(t)$ and $\vec{j}_{\rm orb}(t)$, we can obtain the temporary orbital eccentricity
%\begin{equation} \label{eq:e_orb}
%    e_{\rm orb}(t) = \sqrt{1+\frac{2\vec{j}^2_{\rm orb}\epsilon_{\rm orb}}{G^2 (M_{\rm h}+M_{\rm rem})^2}}
%\end{equation}
%and the pericenter radius 
%\begin{equation} \label{eq:Rp_t}
%    R_{\rm p}(t) = \frac{\vec{j}_{\rm orb}^2}{G(M_{\rm h}+M_{\rm rem})(1+e_{\rm orb})}
%\end{equation}

At the end of the simulations, some simulations have remnant orbits bound to the MBHs ($\epsilon_{\rm orb}(t\to \infty) < 0$), if these remnants continue along their eccentric orbits without other perturbations, they will return to pericenter and be tidally disrupted again. The orbital semi-major axis can be calculated by
\begin{equation} \label{eq:a_orb}
   a_{\rm orb} = -\frac{G(M_{\rm h}+M_{\rm rem})}{2 \epsilon_{\rm orb}(t\to \infty)},
\end{equation}
and the orbital period is set by
\begin{equation} \label{eq:P_orb}
    P_{\rm orb} = 2\pi \sqrt{\frac{a_{\rm orb}^3}{G(M_{\rm h}+M_{\rm rem})}}.
\end{equation}

We also find that the orbital angular momentum of the remnant changes only slightly for bound cases (see Figure \ref{fig:Eorb_Rp_dM}), and the pericenter radius changes negligibly, by only $\sim 10^{-6} R_{\rm p}$. Therefore, the remnant will return to the same pericenter if it moves along the orbit without any perturbations.

Conversely, if the remnants' orbit is unbound ($\epsilon_{\rm orb}(t\to \infty) > 0$), it will be ejected from the system with a kick velocity $v_{\rm kick} = \sqrt{2 \epsilon_{\rm orb}(t\to \infty)}$. If the kick velocity is high enough, the remnant may even escape the galaxy hosting the TDE \citep{manukian_turbovelocity_2013,Gafton_relativistic_2015} and become an intergalactic object.

\begin{figure}		
\centering
 \includegraphics[scale=0.5]{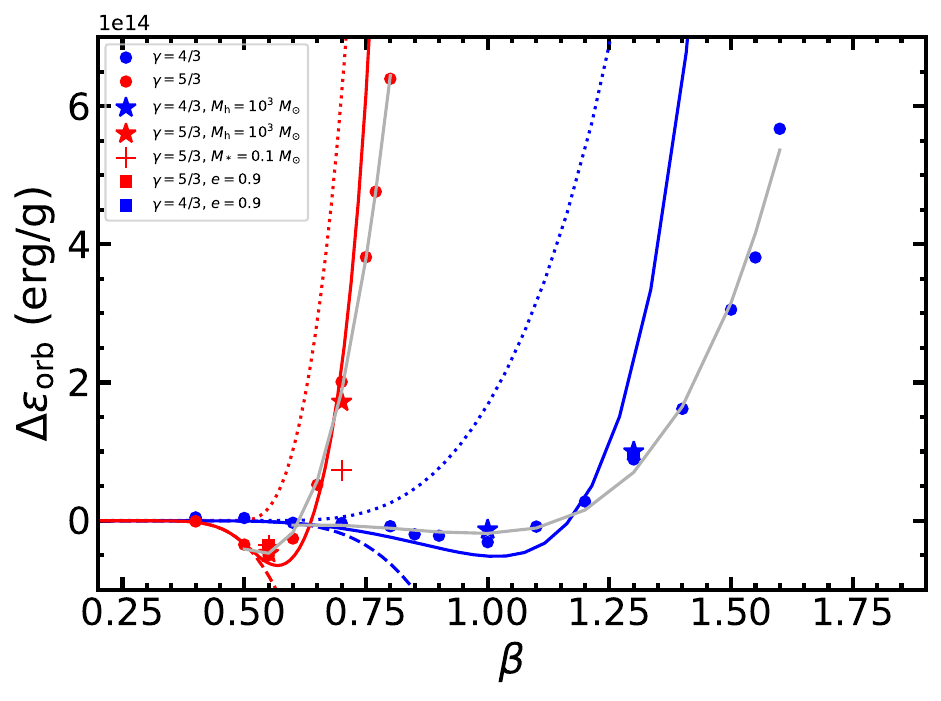}
\caption{Change in the specific orbital energy of the remnant $\epsilon_{\rm orb}({\rm end}) - \epsilon_{\rm orb}(0)$ after a pTDE as a function of $\beta$, for the two types of stellar structures. The data points are obtained at the end of the simulations, with the orbital energy of the remnant given by Eq. (\ref{eq:E_orb}). The dashed, dotted, and solid blue/red lines represent the analytical calculation of the energy deposition to oscillations $\epsilon_{\rm osc}$ (Eq. \ref{eq:E_osc}), the energy gain due to asymmetric mass loss $\epsilon_{\rm ml}$ (Eq. \ref{eq:E_ml}), and the orbital energy change $\epsilon_{\rm ml} - \epsilon_{\rm osc}$ (Eq. \ref{eq:dE_orb}), respectively. When $\beta$ is small, $\epsilon_{\rm osc}$ dominates, leading to the remnant losing specific orbital energy; while for large $\beta$, $\epsilon_{\rm ml}$ dominated, resulting in the remnant gaining specific orbital energy. Different types of the data points depict various parameter setup, the same with that in Figure \ref{fig:dm_beta}. The grey lines represent the fitting result of Eq. (\ref{eq:de_inter}).}
\label{fig:Eorb_beta}
\end{figure}

In Figure \ref{fig:Eorb_beta}, we plot the specific energy of the remnant at the end of the simulations. One can see that the trends for the two stellar structures are similar. 
%When the pericenter is far from the MBH, the tidal force has a negligible effect on the star. 
For small value of $\beta$, the remnant loses specific orbital energy. As $\beta$ increases, the remnant eventually gains specific orbital energy with the energy change increasing with $\beta$. Therefore, if the star is initially in a parabolic orbit, after the partial disruption, the remnant can remain bound or become unbound to the MBH depending on the value of $\beta$.  We define a ``transitional'' impact factor parameter $\beta_{\rm t}$, which is approximately $1.1$ for $\gamma = 4/3$ and $0.62$ for $\gamma=5/3$ polytropic stars. When $\beta \lesssim \beta_{\rm t}$ (or $\beta \gtrsim \beta_{\rm t}$), the remnant is bound (or unbound). We also find that the MBH mass and stellar orbital eccentricity barely affect this orbital energy change.

We interpolate the relation between the specific energy change of the remnant and $\beta$ as
\begin{equation} \label{eq:de_inter}
    \footnotesize{
    \begin{split}
    \Delta \epsilon_{\rm orb} &= \begin{cases}
    7.9 \beta^4 - 19.2 \beta^3 + 15.1 \beta^2 - 3.94 \beta + 0.0035, \gamma=4/3 \\
    121 \beta^3 - 128 \beta^2 + 32.8 \beta + 0.0011, \gamma=5/3
    \end{cases} \\
    &\times 10^{14} {\rm erg \ g^{-1}},
    \end{split}
    }
\end{equation}
which is valid with $0.7 \lesssim \beta \lesssim 1.6$ for $\gamma = 4/3$ and $0.5 \lesssim \beta \lesssim 0.8$ for $\gamma = 5/3$. These fitting results are plotted in Figure \ref{fig:Eorb_beta}, comparing with the simulation results.

\begin{figure}		
\centering
 \includegraphics[scale=0.5]{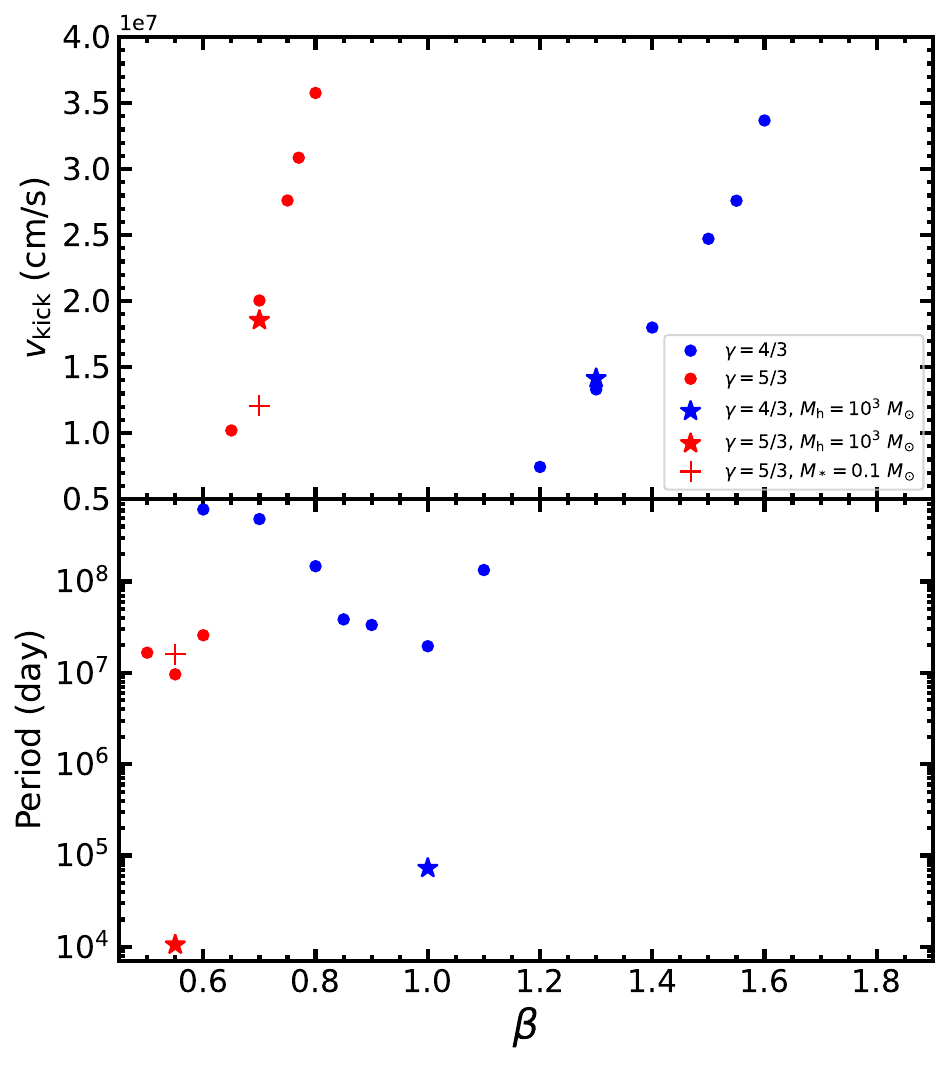}
\caption{Upper and lower panels depict the predicted kick velocity $v_{\rm kick}$ of unbound remnants and the orbital period $P_{\rm orb}$ of bound remnants after pTDEs, respectively. The circular points represent the results of the partial disruption of a polytropic star ($M_* = M_{\odot}$, $R_*=R_{\odot}$) by a $10^6\ M_{\odot}$ mass MBH in an initial parabolic orbit ($e=1$). The star points correspond to the results of pTDEs with IMBHs, while the cross points denote the results of pTDEs with small stars. As shown in Figure \ref{fig:Eorb_beta}, for cases with small $\beta$, the remnant loses orbital energy and becomes bound to the MBH, whereas for cases withe large $\beta$, the remnant is unbound and can fly away with a velocity of $v_{\rm kick}$. In the IMBH cases, the kick velocity exhibits only slight differences from the SMBH cases. However, if the remnant is bound to the IMBH, the predicted period is much shorter.}
\label{fig:vkick_period}
\end{figure}

We also calculate their kick velocity $v_{\rm kick}$ for the unbound remnants and the orbital period $P_{\rm orb}$ for the bound ones, respectively, and plot those in Figure \ref{fig:vkick_period}. $v_{\rm kick}$ can reach up to few $\times 10^7\ {\rm cm/s}$, which is consistent with the results in \cite{manukian_turbovelocity_2013} and \cite{Gafton_relativistic_2015}.  For the bound cases with an SMBH ($10^6\ M_{\odot}$) in a initial parabolic orbit, the orbital period of the remnant is very long. However, if the central BH is an IMBH, the orbital period would be much shorter ($\sim 10$ yrs). Therefore, if the initial orbit of the star is parabolic, we likely cannot detect the recurrence of pTDEs with an SMBH using the current transient surveys, but we could detect those with an IMBH.

In the above analysis, we consider only the isolated system consisting of the black hole and the remnant. However, the stellar distribution around the black hole also plays an important role in determining the subsequent evolution of the remnant's orbit. For unbound remnants with a kick velocity $v_{\rm kick} \simeq {\rm few} \times 10^2\ {\rm km/s}$, they can easily escape the SMBH's influence radius $r_{\rm inf} = GM_{\rm h}/\sigma^2 \simeq 0.5 - 1$ pc, where the velocity dispersion in the galaxy bulge is $\sim 60-90\ {\rm km/s}$ \citep{kormendy_coevolution_2013}. However, if the potential beyond $r_{\rm inf}$ is logarithmic, the remnant will only reach a maximum distance of a few $r_{\rm inf}$, i.e., $\mathrm{e}^{(v_{\rm kick}/2\sigma)^2} r_{\rm inf} \simeq 10 r_{\rm inf}$, meaning it would still remain trapped within the bulge region \citep{Ryu_tidal3_2020}.

For remnant in nearly parabolic or eccentric orbit, they could return to the black hole. However, if the apocenter is comparable to or larger than $r_{\rm inf}$, the orbital period  calculated by Keplerian assumption, i.e., Eq. (\ref{eq:P_orb_bound}) and Figure \ref{fig:vkick_period}, would be influenced by the stellar distribution. The apocenter radius is $\sim 0.4 - 2$ pc for period of $10^7 - 10^8$ day, which is comparable to $r_{\rm inf}$. As a result, the orbital period would be slightly shortened due to the additional potential contributed by other stars.

\subsubsection{The composition of Energy}
\label{subsubsec:energy_relation}

To delve into the underlying physical mechanisms governing the evolution of the remnant's orbit, we analyze how different types of the energy transfer and distribution occur in simulations.

Firstly, we calculate orbital energy of a given cell in the center of mass (COM) reference frame using the equation 
\begin{equation} \label{eq:dE_i}
    dE_i =  \frac12 (\vec{v}_i - \vec{v}_{\rm cm})^2 dm_i - \frac{GM_{\rm h}dm_i}{R_i}.
\end{equation}
Here the COM is situated very close to the MBH due to $M_{\rm h} \gg M_*$, although there might have slight differences if the MBH mass is smaller. $\vec{v}_{\rm cm}$ is the velocity of the COM in a given reference frame, and $R_i$ represents the distance between the COM and a given cell.

We partition the gas into three different parts: the remnant, and two debris streams with negative and positive orbital energies \footnote{For the simulations with initial eccentric orbit $e < 1$, we utilize the initial orbital energy $\epsilon = -GM_{\rm h}(1-e)/(2R_{\rm p})$ to separate these two streams}, as shown in the inset panel of Figure \ref{fig:orbit_sketch}. 

By summing up the orbital energy of the gas in each cell, we obtain the total orbital energy of the remnant gas $E_{\rm orb, rem}$, along with the bound $E_{\rm orb, M1}$ and unbound $E_{\rm orb, M2}$ stripped masses.

\begin{figure}		
\centering
 \includegraphics[scale=0.25]{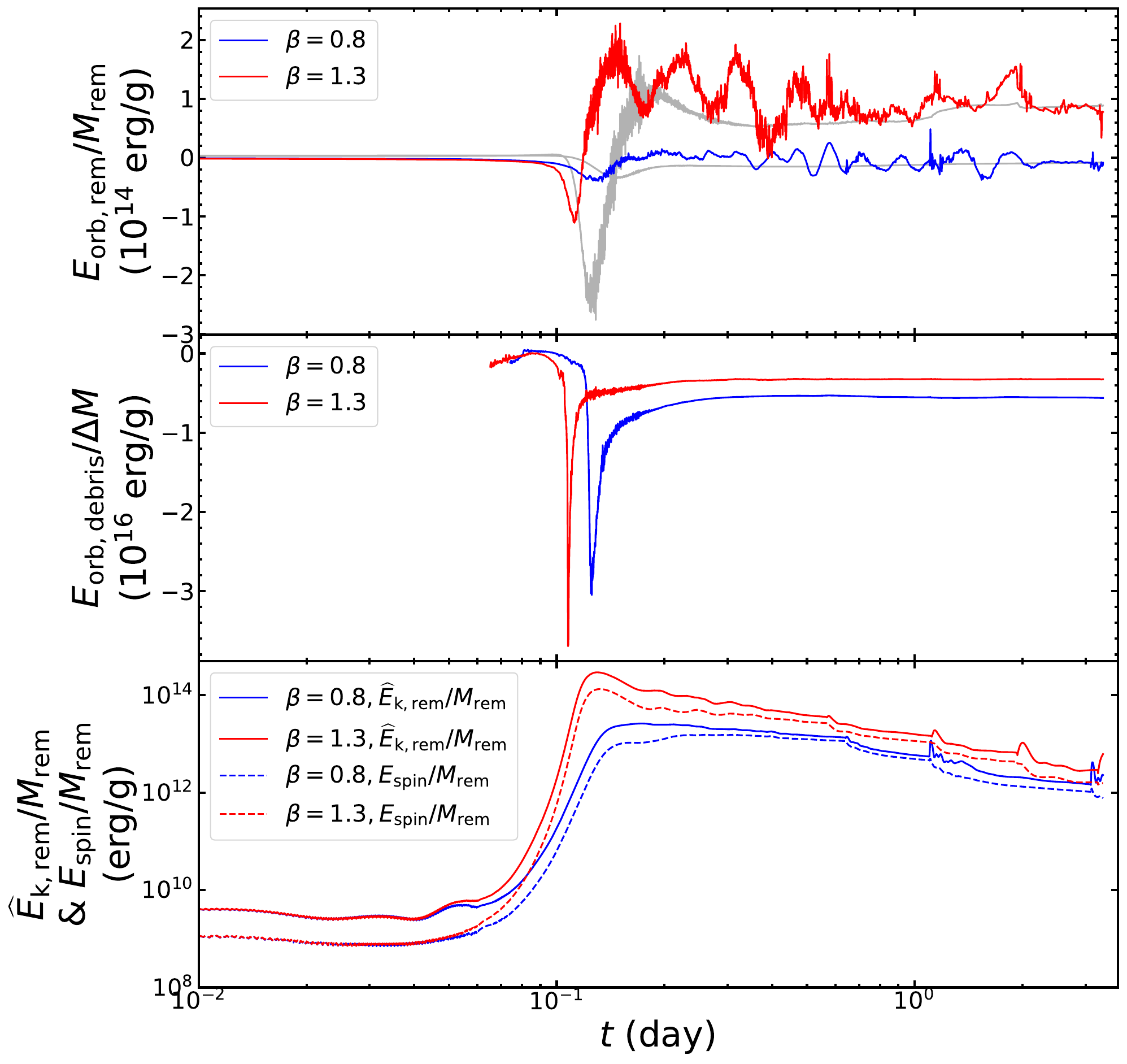}
\caption{Upper, middle and lower panels illustrate the integral specific energy of the remnant gas, debris, and the kinetic energy in the remnant's frame, respectively. The blue and red lines represent the results of partial disruption of a polytropic star ($M_* = M_{\odot}$, $R_*=R_{\odot}$, $\gamma=4/3$) by a $10^6\ M_{\odot}$ mass MBH in an initial parabolic orbit ($e=1$) with $\beta = 0.8$ and $\beta=1.3$, respectively. In the lower panel, we contrast the gas kinetic energy with the rotational energy, derived from Eq. (\ref{eq:Espin}). The gray lines in upper panel depict the specific energy of the remnant, calculated by Eq. (\ref{eq:E_orb}), and exhibit trends akin to the integral orbital energy of the remnant. For the integral orbital energy of the debris, we only display results after the periapsis encounter, as $\Delta M \sim 0$ before the encounter.}
\label{fig:energy_relation}
\end{figure}

In Figure \ref{fig:energy_relation}, we illustrate the evolution of orbital energy of the remnant gas $E_{\rm orb, rem}$ and the debris $E_{\rm orb, debris} = E_{\rm orb, M1}+E_{\rm orb, M2}$. When the star passes by the pericenter, the tidal force deposits some orbital energy into oscillatory modes, leading to a sudden drop in both.

The oscillation causes the inner motion of the remnant. We depict the kinetic energy of the remnant in Figure \ref{fig:energy_relation}:
\begin{equation} \label{eq:Ekk_rem}
    \widehat{E}_{\rm k, rem} = 0.5 \sum [(\vec{v}_i - \vec{v}_0)^2dm_i],
\end{equation}
which is calculated in the remnant's reference frame.

The motion of the gas within the remnant is not entirely chaotic. Some parts of the gas deep inside the remnant oscillate, causing the fluctuation in the remnant's density, which will eventually dissipate, as illustrated in \cite{sharma_partial_2024}. Generally, the gas velocity is azimuthal, as shown in Figure \ref{fig:spin_snapshot}. This is because the tidal torque causes rotation of the star, aligning it with the orbital angular momentum.

The rotational energy of the remnant is define as
\begin{equation} \label{eq:Espin}
    E_{\rm spin, rem} = \frac12 \sum[(\vec{v}_{\rm i}-\vec{v}_0)_{\phi}^2 dm_i],
\end{equation}
where the subscript $\phi$ denotes the velocity component in the azimuthal direction. As illustrated in Figure \ref{fig:energy_relation}, most of the kinetic energy of the remnant comprises rotational energy. 

The stellar spin can influence the subsequent disruption processes if the remnant returns to the pericenter, causing material to fall back sooner and with a higher mass fallback rate \citep{golightly_tidal_2019}. Conversely, if the remnant is ejected, with fast rotation and difference from the normal stars \citep{sharma_partial_2024}, could be potentially observed. We will further discuss the rotation and structure of the remnant in Section \ref{subsubsec:structure}.

\begin{figure}		
\centering
 \includegraphics[scale=0.25]{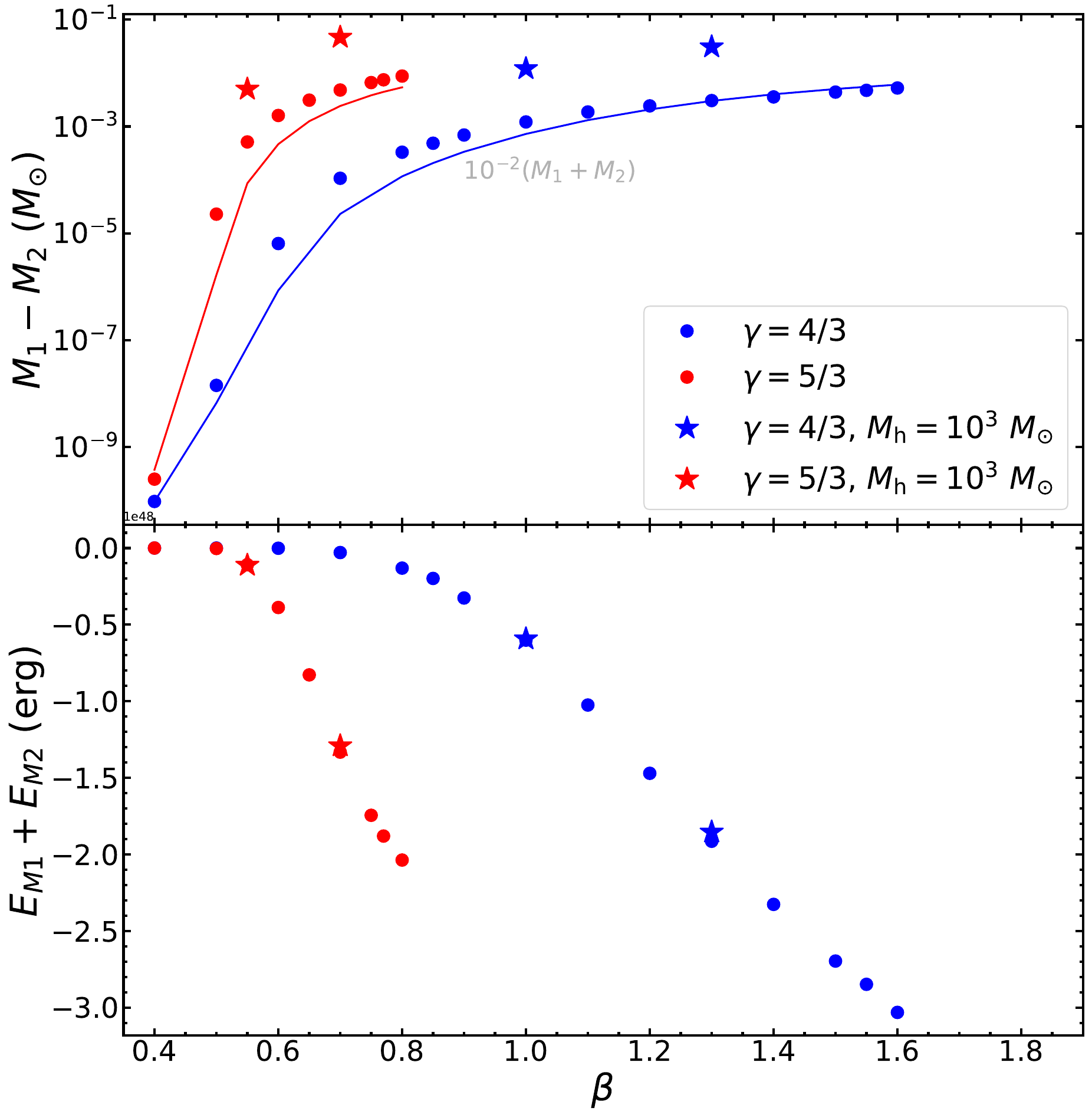}
\caption{Signatures of asymmetric mass loss are depicted in the upper and lower panels. These panels respectively illustrate the mass difference and the total orbital energy of the debris in the simulations with default setup ($M_{\rm h}=10^6\ M_{\odot}$, $M_* = M_{\odot}$, $e=1$) and IMBHs ($10^3\ M_{\rm h}$). Different colors represent different $\gamma$. In the upper panel, solid lines represent the fitting formula of $10^{-2}(M_1+M_2)$, we will discuss the physical reason in section \ref{subsec:osc_strip}. Due to the asymmetric mass loss, the mass difference between the two streams increases, and the total orbital energy of the debris is negative and becomes smaller for larger $\beta$. This elucidates why the remnant can acquire orbital energy in cases with large $\beta$.}
\label{fig:M1_M2}
\end{figure}

To quantitatively analyze the effect of asymmetric mass loss, we examine the mass and specific energy of the debris, as shown in Figure \ref{fig:M1_M2}. We find that the $M_1$ stream bound to the MBH, always outweighs the unbound steam, $M_2$, a discrepancy that amplifies with increasing $\beta$. Moreover, for larger $\beta$, the total orbital energy of both streams is negative and becomes smaller, and consequently the asymmetric mass loss effect becomes more significant. This energy distribution mechanism explains why the remnant can become unbound and exhibit higher kick velocity for large $\beta$.

\subsubsection{Remnant's structure}
\label{subsubsec:structure}
After the partial disruption, some material from the stellar surface is tidal stripped away. The remaining remnant undergoes tidal spinning and adiabatic expansion, transforming it into a structure distinct from a normal star. In the following analysis, we will examine the rotation, density, and radius of the remnant.

\begin{figure*}	
\centering
\begin{minipage}[t]{0.48\textwidth}
\centering
 \includegraphics[scale=0.27]{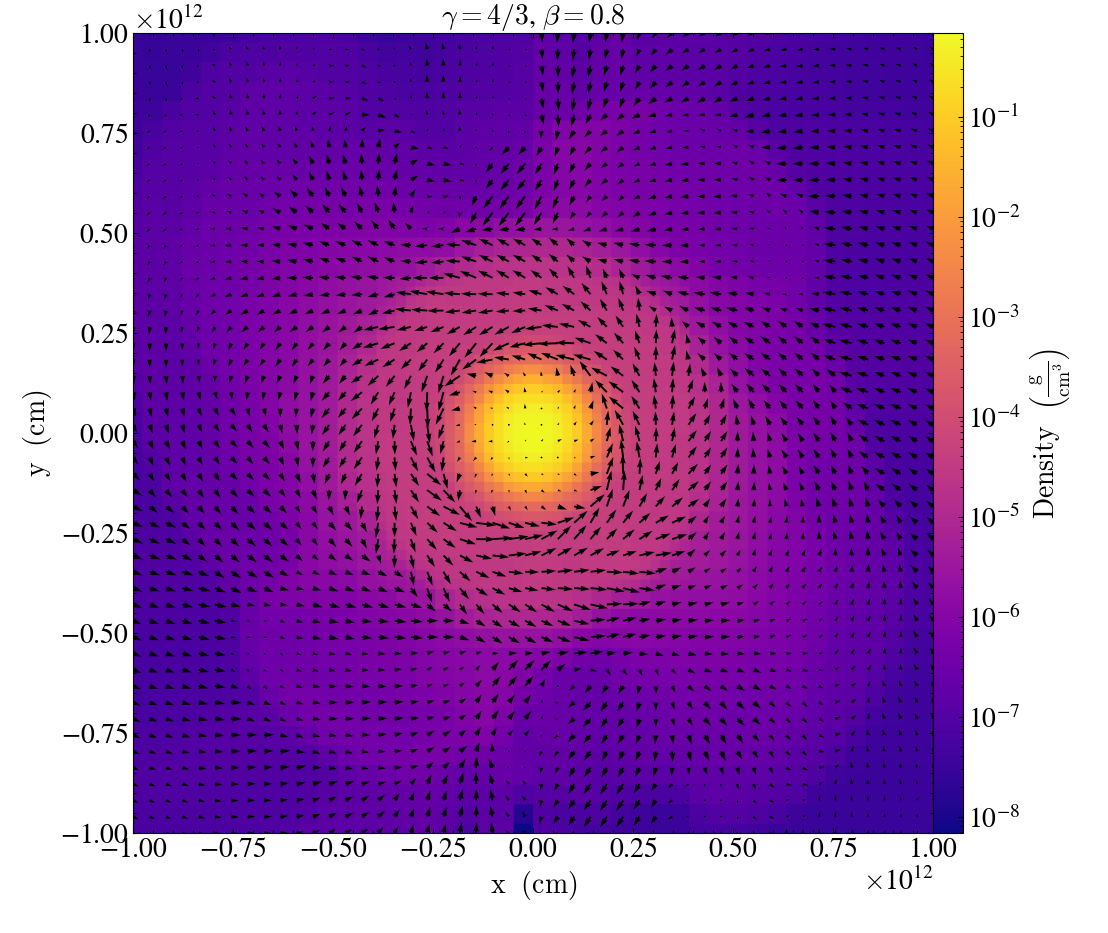}
 \end{minipage}
 \begin{minipage}[t]{0.48\textwidth}
 \centering
 \includegraphics[scale=0.27]{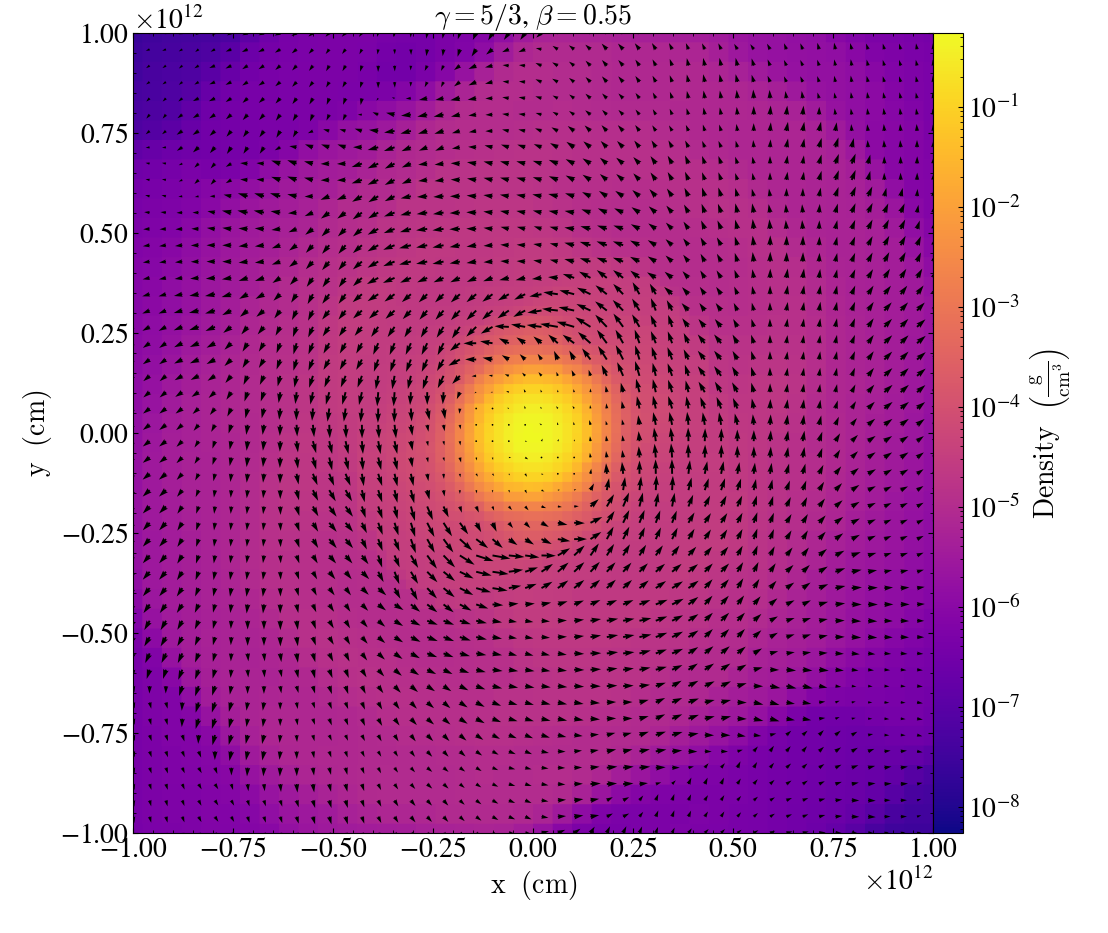}
  \end{minipage}
 \begin{minipage}[t]{0.48\textwidth}
 \centering
 \includegraphics[scale=0.27]{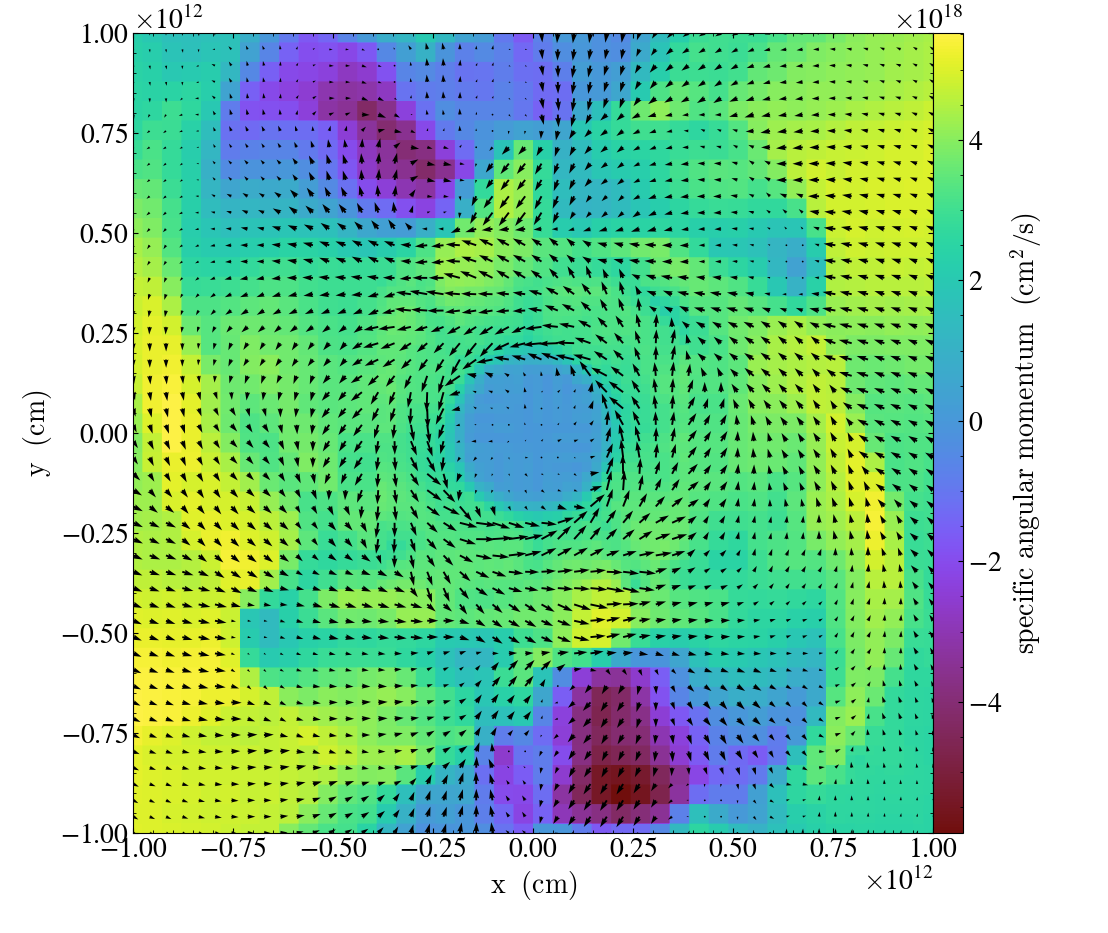}
  \end{minipage}
 \begin{minipage}[t]{0.48\textwidth}
 \centering
 \includegraphics[scale=0.27]{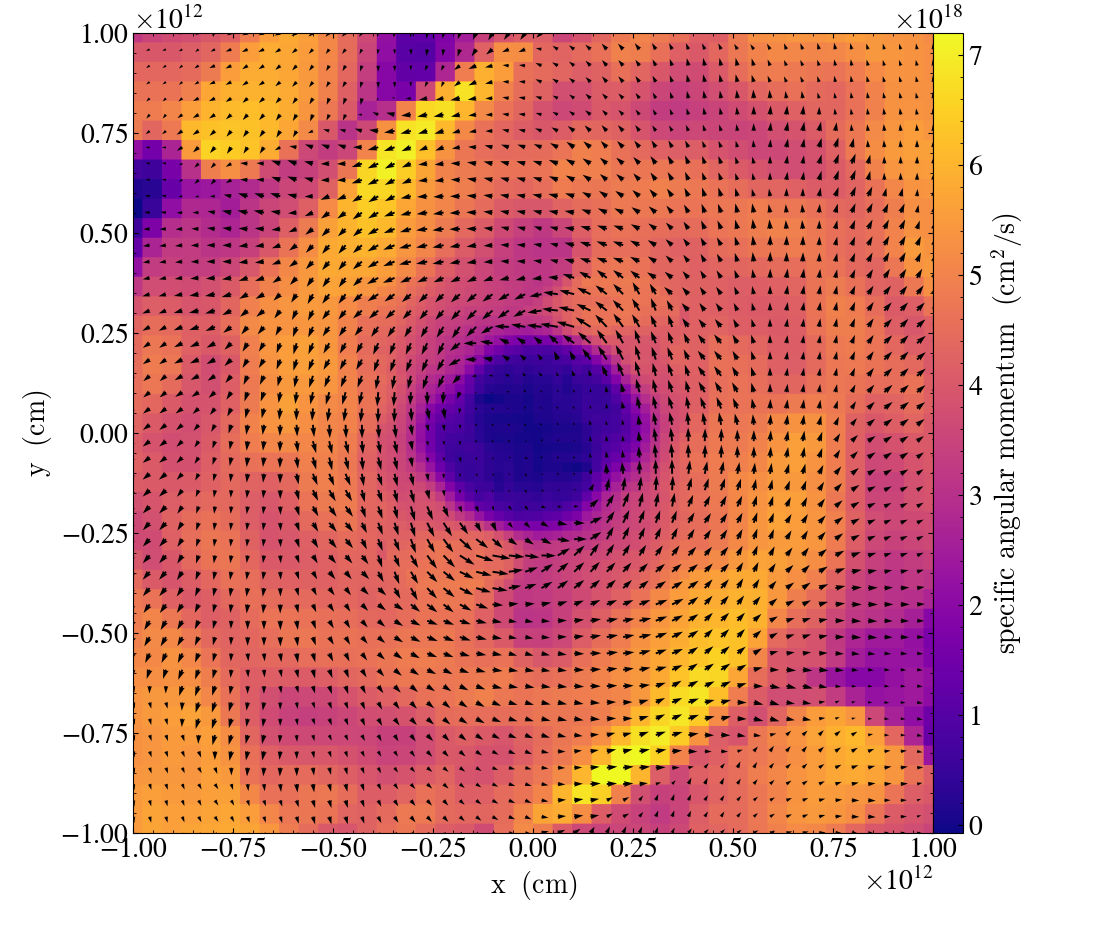}
  \end{minipage}
 \begin{minipage}[t]{0.48\textwidth}
 \centering
 \includegraphics[scale=0.27]{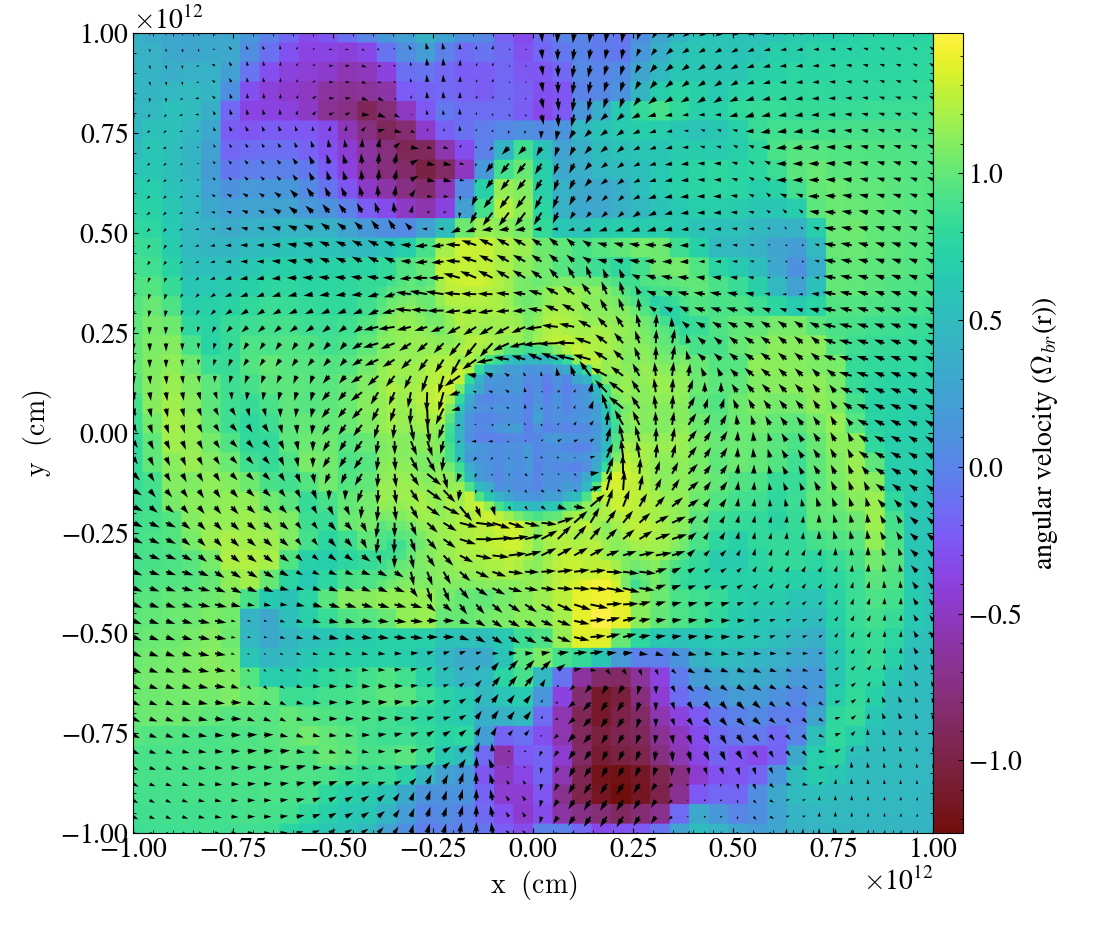}
  \end{minipage}
 \begin{minipage}[t]{0.48\textwidth}
 \centering
 \includegraphics[scale=0.27]{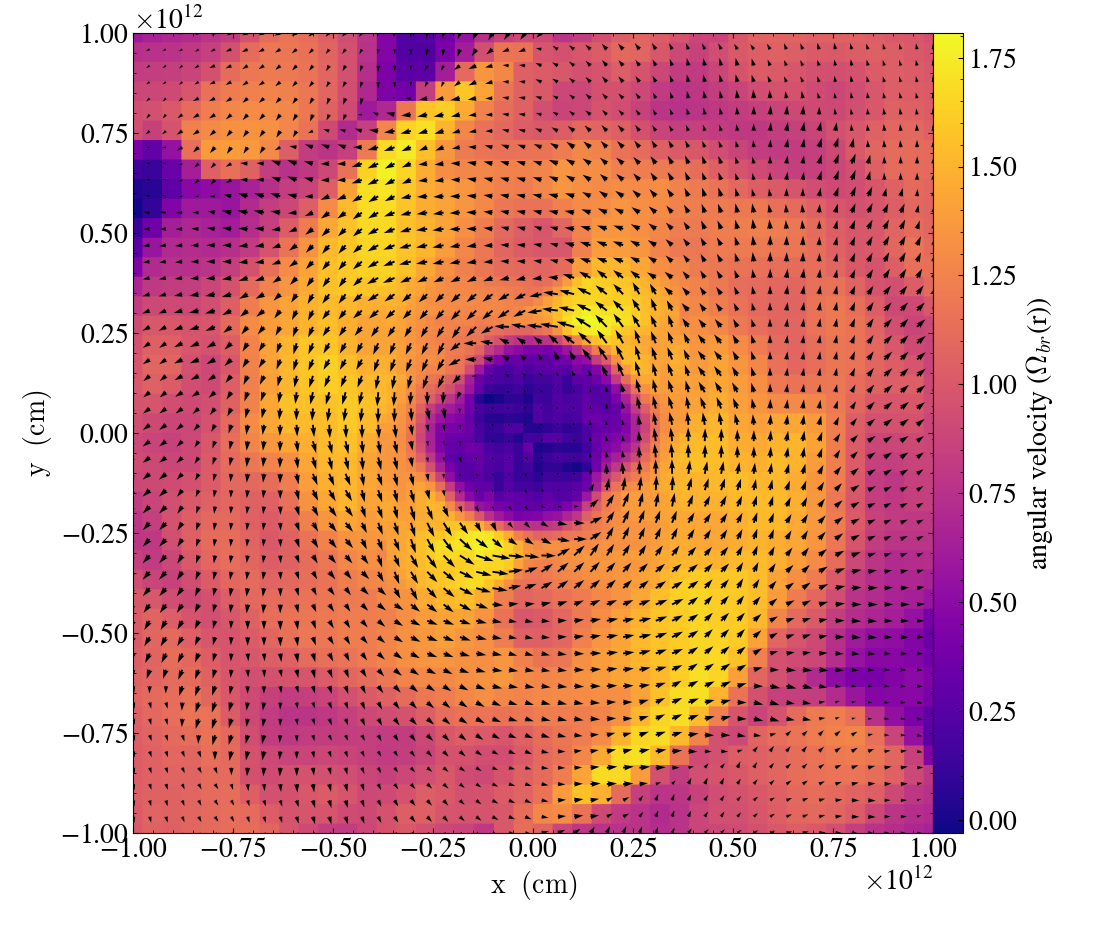}
  \end{minipage}
\caption{Upper, middle, and lower panels depict the density projection, specific angular momentum, and the angular velocity distribution of the remnant on the orbital plane, respectively. These figures illustrate the results of partial disruption of a polytropic star ($M_* = M_{\odot}$, $R_*=R_{\odot}$) by a $10^6\ M_{\odot}$ mass MBH in an initial parabolic orbit ($e=1$) with different $\gamma$ and $\beta$, as indicated in the title. The remnants exhibits a distinct central region surrounded by a thin outer thin envelope. While the central region rotates almost rigidly, the envelope displays differential rotation, approaching the break-up velocity $\Omega_{\rm br}$.}
\label{fig:spin_snapshot}
\end{figure*}

Figure \ref{fig:spin_snapshot} displays the density, specific angular momentum $j_{\rm spin} = \Omega_{\rm spin} r^2$, and the angular velocity of rotation $\Omega_{\rm spin}$, where $r$ is the distance from the remnant's center. The remnant exhibits a dense central region surrounded by a diffuse, thin envelope. The central region rotates almost rigidly, and the envelope experiences differential rotation, with a rapid rotation comparable to the break-up velocity $\Omega_{\rm br} = \sqrt{GM(r)/r^3}$, where $M(r)$ is the mass enclosed at $r$. These results are close to those of \cite{sharma_partial_2024}. The morphology of the six G-objects detected in our galactic center are similar to that of the remnant \citep{ciurlo_population_2020}.

\begin{figure}	
\centering
\includegraphics[scale=0.25]{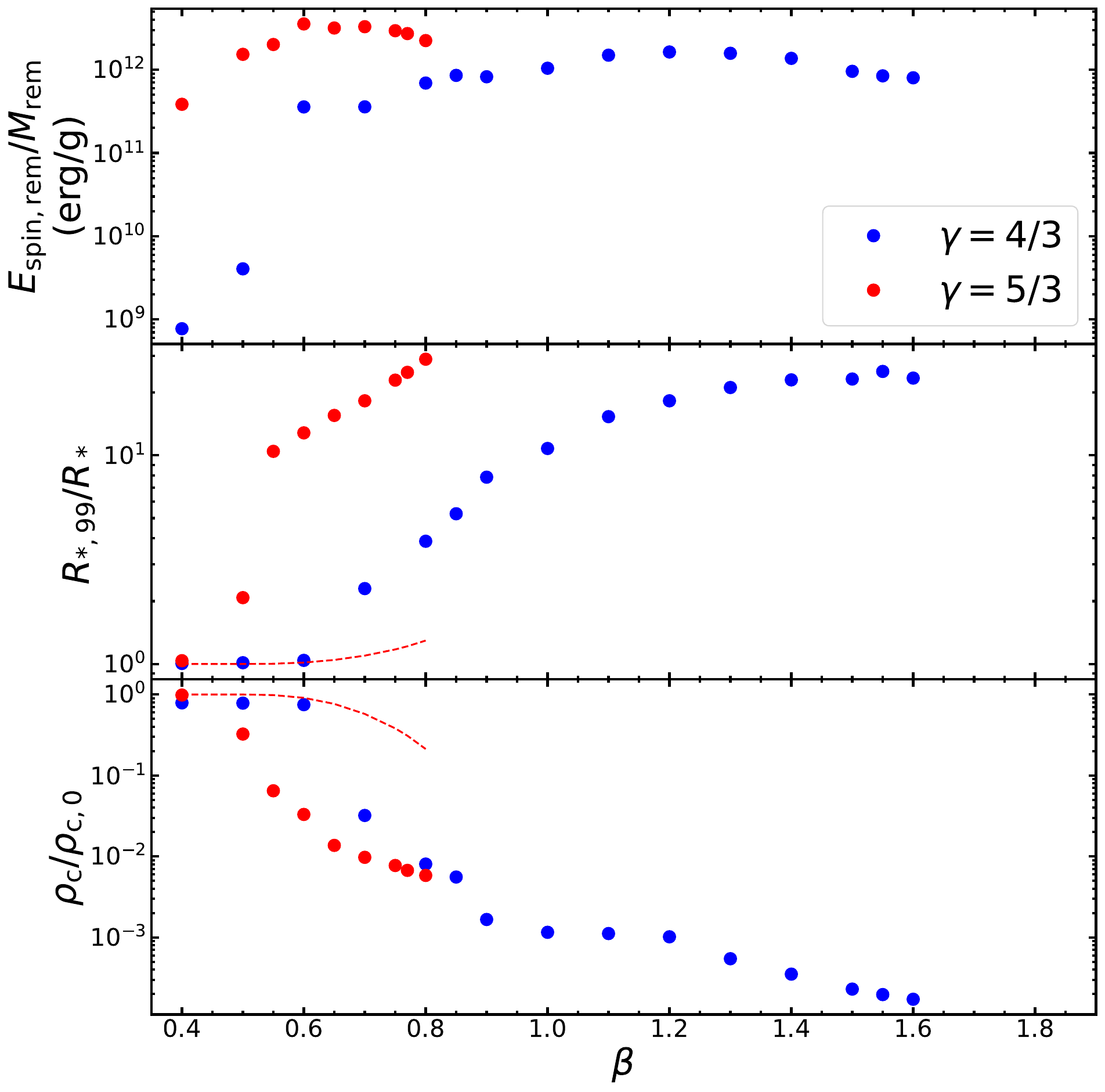}
\caption{Remnant properties are illustrated in the upper, middle, and lower panels. These panels respectively illustrate the specific rotational energy, the ratio of the remnant's radius to the initial stellar radius, and the ratio of the remnant's central density to the initial stellar central density in the simulations with default setup ($M_{\rm h}=10^6\ M_{\odot}$, $M_* = M_{\odot}$, $R_* = R_{\odot}$, $e=1$). The average radius of the remnant, denoted as $R_{*, 99}$, is defined as the radius enclosing $99\%$ of the remnant's mass. The dashed lines in middle and lower panels represent the expected result of an adiabatic mass-losing star.}
\label{fig:remnant_property}
\end{figure}

The radius of the central region expands to several times the initial stellar radius, whereas the envelope reaches a scale exceeding to $\gtrsim 10$ times the initial radius. We estimate the average remnant radius using $R_{*,99}$ encompassing $99 \%$ of the remnant's mass. Figure \ref{fig:remnant_property} illustrates $R_{*, 99}$ and the central density of the remnant at the end of each simulation.

When the $\beta$ is small, the specific rotational energy increases with $\beta$, whereas it stabilizes at a nearly constant value for the large $\beta$. Meanwhile, the remnant's radius continuous to increase, and the central density decreases. Consequently, the remnant is expected to become more susceptible to be disrupted as it returns to the pericenter. In cases with small $\beta$, the central density of the remnant only changes slightly, indicate that they would undergo multiple pTDEs before experiencing full disruption. In section \ref{sec:pTDE}, we will provide further predictions regarding the evolution of pTDEs, including the number of encounters and how the remnant's orbit evolves during each encounter.

Determining the radius of the remnant, $R_{*, 99}$, poses challenges and is tricky, particularly as most of the region comprises the thin gas envelope that is not strictly spherical. We anticipate that upon the remnant's return to the pericenter, much of the thin envelope would be tidally stripped away. However, since our work only simulates a single pTDE process, further investigation into multiple passages is warranted in future studies.

%%%%%%%%%%%%%%%%%%%%%%%%%%%%%%%%%%%%%%%%%
\section{Analysis of the remnant orbital change} \label{sec:bound_unbound}

\subsection{Oscillation and asymmetric mass loss}
\label{subsec:osc_strip}
When the star approaches the central MBH near the pericenter, the tidal force induces oscillations in the star, transferring some orbital energy to the star. The physics of adiabatic, non-radial oscillations was thoroughly examined by \cite{Press_osc_1977} and extended by \cite{Lee_osc_1986}, who corrected an error in the original Press-Teukolsky paper.

Here we briefly recapitulate their calculations here. The specific energy deposition into oscillations can be expressed as a product of dimensional quantities and a dimensionless function.
\begin{equation} \label{eq:E_osc}
    \epsilon_{\rm osc} = \left(\frac{GM_*}{R_*}\right) \left(\frac{M_{\rm h}}{M_*}\right)^2 \sum\limits_{l=2,3,...} \left(\frac{R_*}{R_{\rm p}}\right)^{2l+2} T_l,
\end{equation}
where the dimensionless function $T_l$ depends on the stellar structure, the corresponding oscillatory modes, and $\beta$.

We use the GYRE stellar oscillation code to numerically calculate the oscillation of polytropic stars with $\gamma=5/3$ and $4/3$ \citep{townsend_gyre_2013}. Then we use Eq. (2.4-2.7) of \cite{Lee_osc_1986} to calculate $T_l$.

\begin{figure}		
\centering
 \includegraphics[scale=0.5]{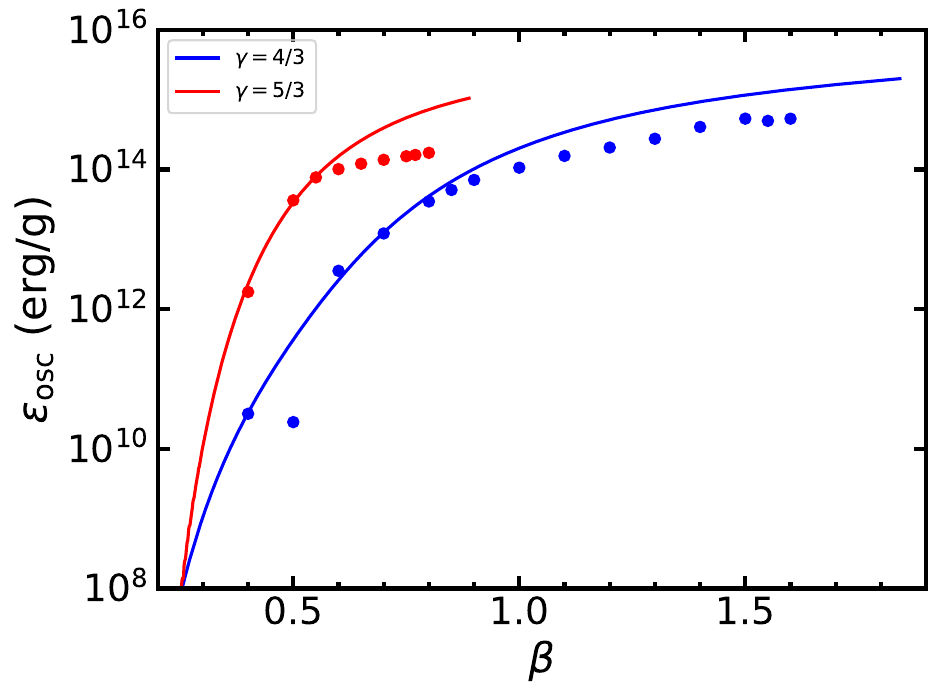}
\caption{Deposition of energy into oscillatory modes during the partial disruption of a polytropic star ($M_* = M_{\odot}$, $R_*=R_{\odot}$) by a $10^6\ M_{\odot}$ mass MBH in an initial parabolic orbit ($e=1$). The solid lines represent the analytical prediction given by Eq. (\ref{eq:E_osc}). The data points depict the energy deposition near the pericenter estimated by the drop of the orbital energy in the simulation (refer to Figure \ref{fig:Eorb_Rp_dM}). The simulation results are consistent with the analytical prediction, except for the cases with large $\beta$, where the linear approximation is no longer applicable.}
\label{fig:Eosc}
\end{figure}

To quantitatively analyze this effect in our simulations, we compute the orbital energy deposition near the pericenter and compare it with the analytical Eq. (\ref{eq:E_osc}), as shown in Figure \ref{fig:Eosc}. We find consistency between the two, except for large $\beta$ close to $\beta_c$. 

For $\beta \gtrsim 0.5$, the star loses mass due to tidal force, resulting in two debris steams -- one is bound to the MBH and the other is unbound. Because the bound stream experiences more intense tidal effects, the mass loss is asymmetric that causes the bound one has more mass. This asymmetry can be attributed to the differences in the tidal field, which we quantify in the following analysis.

The total orbital energy of two debris stream can be expressed as
\begin{equation} \label{eq:EM1_EM2}
    E_{\rm M1} + E_{\rm M2} \sim - \frac{GM_{\rm h}M_1 R_*}{(R_{\rm t}-R_*)^2} + \frac{GM_{\rm h}M_2 R_*}{(R_{\rm t} + R_*)^2}.
\end{equation}
Here, we replace $R_{\rm p}$ with $R_{\rm t}$ to align with simulation results. Since $R_* \ll R_{\rm t}$, this equation can be approximated using a first-order expansion with respect to $R_{\rm t}$, i.e., $1/(R_{\rm t} \pm R_*)^2 \simeq 1/R_{\rm t}^{2} \mp 2R_*/R_{\rm t}^3$, yielding 
\begin{equation} \label{eq:EM1_EM2_2}
    E_{\rm M1} + E_{\rm M2} \sim - \frac{GM_{\rm h} (M_1-M_2)R_*}{R_{\rm t}^2}-2\frac{GM_{\rm h}R_*^2 \Delta M R_*^2}{R_{\rm t}^3}.
\end{equation}
This energy loss from the debris is transferred to the remnant, leading to a specific energy gain for the remnant given by  $\epsilon_{\rm ml} \simeq -(E_{\rm M1} + E_{\rm M2})/M_{\rm rem}$, or
\begin{equation} \label{eq:E_ml}
\begin{split}
    \epsilon_{\rm ml} &\simeq 2\frac{GM_{\rm h}R_*^2}{R_{\rm t}^3}\frac{\Delta M}{M_{\rm rem}} + \frac{GM_{\rm h} R_*}{R_{\rm t}^2}\frac{M_1-M_2}{M_{\rm rem}}\\
    &\simeq 2\frac{GM_*}{R_*} \frac{\Delta M}{M_{\rm rem}} +\frac{GM_{\rm h} R_*}{R_{\rm t}^2}\frac{M_1-M_2}{M_{\rm rem}}.
\end{split}
\end{equation}
These two terms correspond to the distance difference to the MBH and mass difference between the two streams, respectively, and are the primary causes of the asymmetry in mass loss. 

Previous studies \citep{Stone_Compression_2013,metzger_luminous_2022,kremer_hydrodynamics_2022}
have typically considered only the first term. Since both terms should be comparable, we estimate the total energy gain as
\begin{equation} \label{eq:E_ml2}
    \epsilon_{\rm ml} \simeq f_{\rm ml} \frac{GM_*}{R_*} \frac{\Delta M}{M_{\rm rem}},
\end{equation}
where $f_{\rm ml}$ is a free parameter that is expected to be close to unity.

\begin{figure}		
\centering
 \includegraphics[scale=0.5]{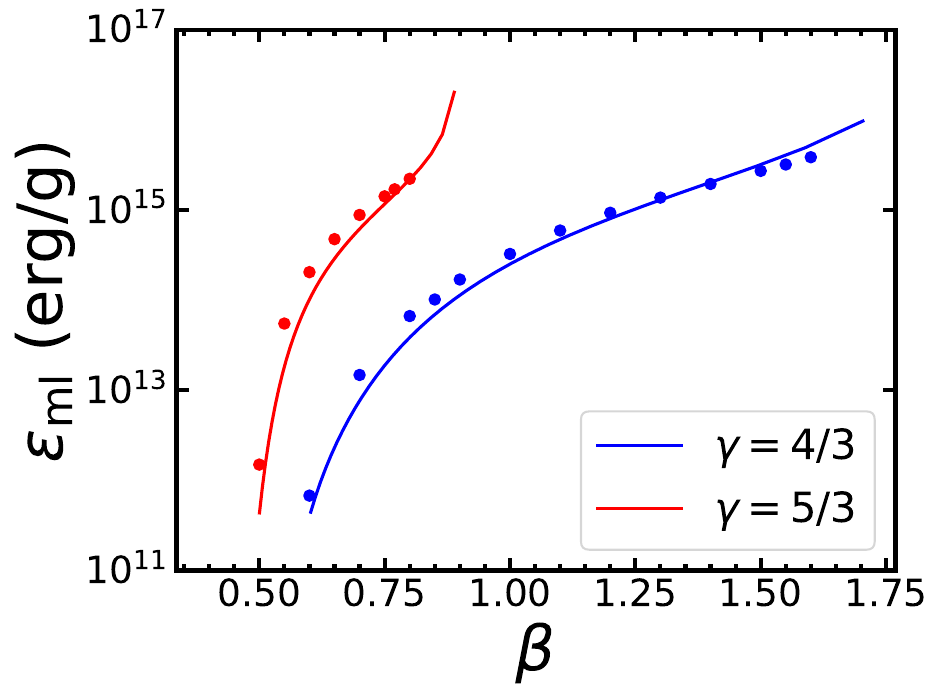}
\caption{Energy gain of the remnant caused by asymmetric mass loss in partial disruption of a polytropic star ($M_* = M_{\odot}$, $R_*=R_{\odot}$) by a $10^6\ M_{\odot}$ mass MBH in an initial parabolic orbit ($e=1$). The data points represent the energy gain of the remnant estimated by $E_{\rm orb, debris}/M_{\rm rem}$. The solid lines represent the analytical estimates given by Eq. (\ref{eq:E_ml2}), where $f_{\rm ml} \sim 1$ and 1.5 for $\gamma = 5/3$ and $4/3$, respectively. Overall, the simulation results are generally consistent with the analytical prediction.}
\label{fig:E_ml}
\end{figure}

In Figure \ref{fig:E_ml}, we compare this analytical estimate with the simulation results, obtained by $E_{\rm orb, debris}/M_{\rm rem}$. We find good agreement between them ($f_{\rm ml} \simeq 1$ and $1.5$ for $\gamma = 5/3$ and $4/3$, respectively). Interestingly, Eq. (\ref{eq:E_ml2}) is independent of the BH mass, as found in previous work \cite{manukian_turbovelocity_2013}. However, \cite{Gafton_relativistic_2015} found that $\epsilon_{\rm ml}$ would be larger for large BH mass when considering the general relativity effects.

Next, we will estimate the mass difference of the streams, $M_1-M_2$. Since the two terms in Eq. (\ref{eq:E_ml}) are comparable,
we can obtain the mass difference as
\begin{equation} \label{eq:M1-M2}
    M_1 - M_2 \sim \left(\frac{M_{\rm h}}{M_*}\right)^{-1/3}\Delta M_*.
\end{equation}
This estimate is consistent with the simulation results (see Figure \ref{fig:M1_M2}).

These two effects, the tidally induced oscillation and asymmetric mass loss, are the physical mechanisms behind the orbital change of the remnant. The change in orbital energy of the remnant is given by
\begin{equation} \label{eq:dE_orb}
    \Delta \epsilon_{\rm orb} \simeq \epsilon_{\rm ml} - \epsilon_{\rm osc},
\end{equation}
We compare it with the simulation results in Figure \ref{fig:Eorb_beta}.

\subsection{Bound vs. unbound remnant: transition point}
\label{subsec:unbound_bound}
When $\beta \lesssim \beta_{\rm t}$, $\epsilon_{\rm osc}$ dominates, it causes the remnant losing orbital energy after the pTDE. Conversely, when $\beta \gtrsim \beta_{\rm t}$, $\epsilon_{\rm ml}$ dominates, it cause the remnant gaining orbital energy. Here we will determine the transitional impact factor $\beta_{\rm t}$ and obtain the parameter dependence of the orbital change.

The deposition energy of $\epsilon_{\rm osc}$ is the summation of each oscillatory modes. Considering that $R_* \ll R_{\rm p}$, $l=2$ mode predominantly contributes. Thus, we have $\epsilon_{\rm osc} \sim \beta^6 (GM_*/R_*) T_2$. By setting $\epsilon{\rm osc} = \epsilon_{\rm ml}$, we obtain $\beta^6 T_2 = \Delta M /(M_*-\Delta M)$. As both sides of this equation depends solely on $\beta$ and stellar structure, $\beta_{\rm t}$ is solely determined by the stellar structure. For the two polytropic stars we consider here, $\beta_{\rm t} \simeq 1.1$ for $\gamma = 4/3$ and $\beta_{\rm t} = 0.62$ for $\gamma = 5/3$.

%\begin{figure}	
%\centering
%\begin{minipage}[t]{0.23\textwidth}
%\centering
% \includegraphics[scale=0.4]{gyre_43.pdf}
% \end{minipage}
% \begin{minipage}[t]{0.23\textwidth}
%\centering
% \includegraphics[scale=0.4]{gyre_53.pdf}
% \end{minipage}
%\caption{Tidal oscillation}
%\label{fig:gyre}
%\end{figure}

\subsection{Parameter dependence for the remnant orbit}
\label{subsec:dependence}
Since $\epsilon_{\rm osc}$ is predominantly influenced by the $l=2$ mode when $M_{\rm h} \gg M_*$, we find
\begin{equation} \label{eq:dE_orb2}
    \Delta \epsilon_{\rm orb} \sim \frac{GM_*}{R_*} \left(\frac{\Delta M}{M_{\rm rem}} - \beta^6 T_2\right),
\end{equation}
which is independent of $M_{\rm h}$. As $\Delta M/M_*$ depends only on $\beta$ and stellar structure, as shown in Eq. (\ref{eq:dm_beta}) and Figure \ref{fig:dm_beta}, $\Delta \epsilon_{\rm orb}$ depends only on stellar properties (mass, radius and structure) and $\beta$. Interestingly, this indicate that the orbital change is independent on $M_{\rm h}$. To validate these relationships, we also simulate pTDEs with IMBH ($10^3\ M_{\odot}$) and smaller stars ($M_* = 0.1\ M_{\odot}$, $R_* = 0.16\ R_{\odot}$), including the eccentric cases ($e = 0.9$). The simulation results in Figure \ref{fig:Eorb_beta} confirm this trend. 

Using Eq. (\ref{eq:dE_orb2}), we can estimate the orbital energy change in various systems. When $\epsilon_{\rm osc}$ dominates, we can estimate the orbital period of the remnant after the encounter for a pTDE on an initial parabolic orbit, i.e., 
\begin{equation} \label{eq:P_orb_bound}
    P_{\rm orb} \simeq 10^7 \left(\frac{M_{\rm h}}{10^6\ M_{\odot}}\right) \left(\frac{M_*}{M_{\odot}}\right)^{-3/2} \left(\frac{R_*}{R_{\odot}}\right)^{3/2}\ {\rm day},
\end{equation}
where we substitute the maximum energy loss of $\epsilon_{\rm osc} \sim 10^{14} (M_*/M_{\odot})/(R_*/R_{\odot})\ {\rm erg\ g^{-1}}$ into it.

\section{Long-term evolution of repeating pTDEs}
\label{sec:pTDE}

If the star has not been fully disrupted or ejected after the periapsis passage, as previously discussed, the remnant will undergo another tidal stripping or disruption upon returning to the pericenter. In this work, we simulate only a single passage. To explore the evolution of repeating pTDEs, we need to extrapolate subsequent processes from our simulations results, considering both the stellar structure and the remnant's orbit.

We begin with a simplified method to calculate the mass loss evolution by assuming that the star loses mass adiabatically near the pericenter, leading to a readjustment of its structure. If the star becomes less centrally concentrated after one encounter, it will be more susceptible to disruption in the next encounter.

In \cite{dai_adiabatic_2013}, the evolution of a mass-losing star’s structure was studied when it overflows its Roche lobe in an extreme mass-ratio circular binary system. We adopt their method to calculate the stellar structure’s evolution.

When the star is tidally stripped after a passage, the remnant will adiabatically readjust to a new equilibrium, maintaining a constant local effective entropy $\widetilde{S}(M) = P^{3/5}/\rho$, where $P$, $\rho$, and $M$ are the local pressure, density, and enclosed mass, respectively.

For a $\gamma = 5/3$ polytropic star, which represents a fully convective star, the effective entropy is uniform throughout the star, i.e., $\widetilde{S}(M) = const$, it remains a $\gamma = 5/3$ polytrope after the mass loss. According to the properties of polytropic stars, the mass-radius relation can be expressed as $R_* \propto \rho_{\rm c}^{-1/6} \propto \bar{\rho}_*^{-1/6} \propto (M_*R_*^{-3})^{-1/6}$, leading to $R_* \propto M_*^{-1/3}$. Therefore, as the star loses mass, its radius will expand, and $\beta = R_{\rm t}/R_{\rm p} \propto R_* M_*^{-1/3} \propto M_*^{-2/3}$ will increase, making the star more vulnerable to disruption. When $\beta \gtrsim 0.9$, the star will be fully disrupted in the next encounter.

\begin{figure}		
\centering
 \includegraphics[scale=0.5]{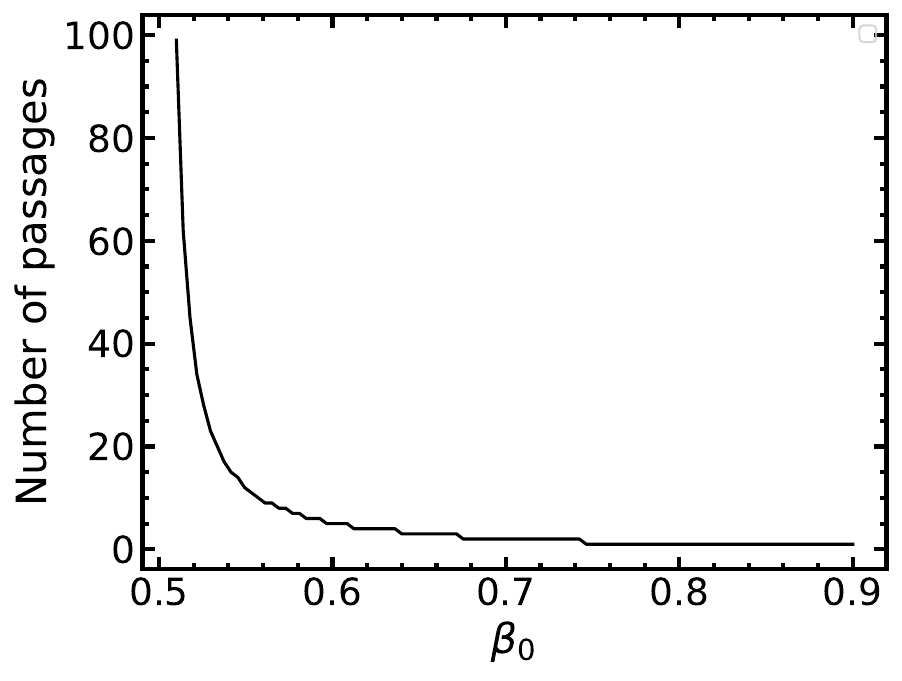}
\caption{Number of passages that a $\gamma = 5/3$ polytrope star undergoes before experiencing full disruption vs. the initial $\beta_0$, which is calculated using the adiabatic mass loss assumption and the orbital change has not taken into account.}
\label{fig:num_adia}
\end{figure}

Using Eq. (\ref{eq:dm_beta}), we can track the mass loss evolution and calculate the total number of passages, as shown in Figure \ref{fig:num_adia}. When $\beta \sim 0.5$, there is almost no mass loss, allowing the star to survive many passages. For $0.5 \lesssim \beta \lesssim 0.6$, the star generally undergoes dozens of passages. For larger $\beta$, only a few passages occur.

Many studies have used the adiabatic mass loss approximation, as we do here, to estimate stellar structure evolution \citep{dai_adiabatic_2013,MacLeod_Spoon_2013,MacLeod_Illuminating_2014,Liu_Tidal_2023}. This approximation has been widely employed in stable mass transfer in binary systems. However, when tidal forces significantly perturb the star and induce oscillations, entropy may no longer remain constant.. 
%We note that the tidal force can deform the star and inject some kinetic energy into the star when the star passes by the pericenter. 
Consequently, the assumption of adiabatic evolution may not hold. Despite this, tidal interactions likely still make the tidally stripped star more vulnerable. As shown in Figure \ref{fig:remnant_property}, the remnant’s density decreases, and its radius expands after the pTDE, making the star more susceptible to disruption if it returns to the pericenter.

After the tidal stripping, the remnant consists of a central region plus a thin outer envelope, no longer resembling a typical star. It is challenging to provide a robust analytical calculation for the stellar structure after multiple passages.
However, we can offer a simplified estimate of the evolution.  If the pericenter radius is smaller than the radius for full disruption, i.e., $R_{\rm p} \lesssim R_{\rm t}/\beta_{\rm c}$, the remnant will be fully disrupted. The critical value for full disruption are $\beta_{\rm c} = 0.9$ and $1.85$ for the polytropic stars with $\gamma = 5/3$ and $4/3$, respectively. Since the radius for full disruption depends on the central density of the star, we allow the radius for full disruption to evolve as $R_{\rm t}/\beta_{\rm c} \propto \rho_{\rm c}^{-1/3}$ after each passage \citep{ryu_tidal_2020}. pTDEs can continue until full disruption ($R_{\rm p} \lesssim R_{\rm t}/\beta_{\rm c}$ is satisfied) or the remnant is ejected. In each passage, we assume $\overline{\rho}_*/\rho_{\rm c}$ remains constant, yielding a new value of $\beta = R_{\rm t}/R_{\rm p}$ to estimate the subsequent changes in the orbit and the remnant. 

\begin{figure}		
\centering
\subfigure{
 \includegraphics[scale=0.5]{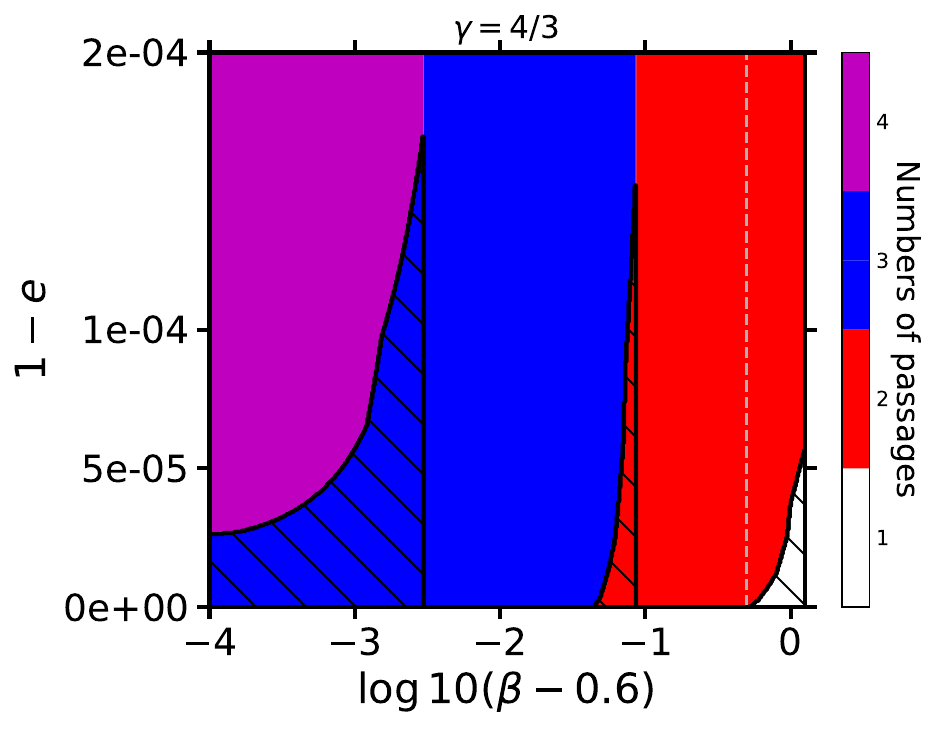}
}
\subfigure{
 \includegraphics[scale=0.5]{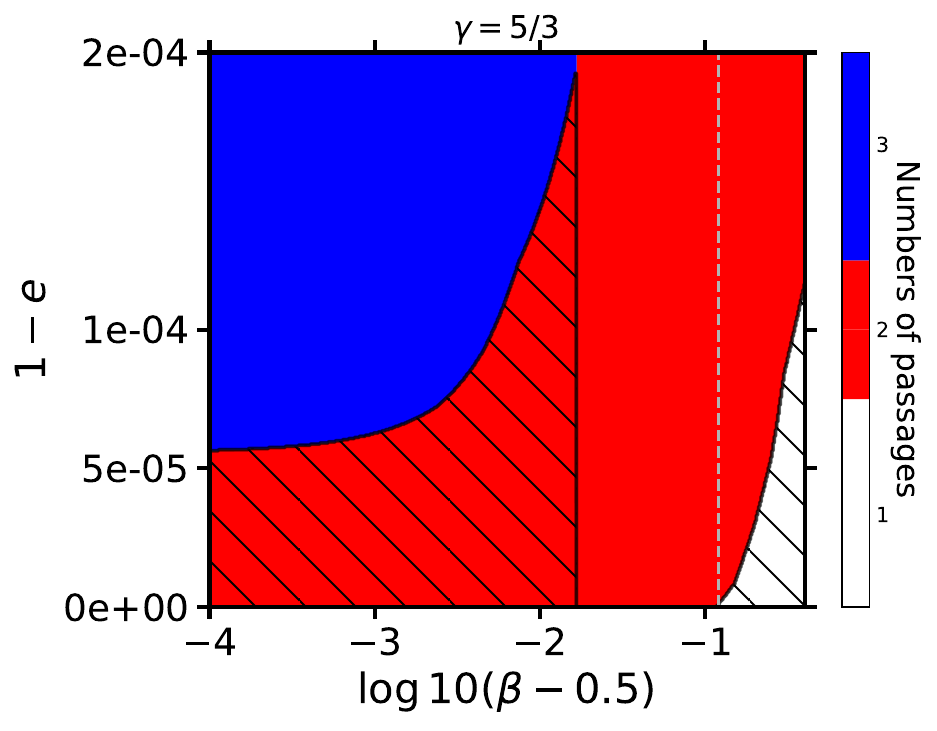}
}
\caption{Number of passages that a star on an initial orbit with $\beta$ and $e$ undergoes  before either being fully disrupted or ejected ($M_* = M_{\odot}$, $R_*=R_{\odot}$ and $M_{\rm h} = 10^6\ M_{\odot}$). The shaded region enclosed by black lines delineate the parameter space where the star is ejected. Outside these region, the star will be fully disrupted after several passages. The grey dashed line represent the transitional impact factor $\beta_{\rm t}$ indicating whether the remnant gains or loses orbital energy after the first passage. For the initial orbit with $e \sim 1$ and $\beta \gtrsim \beta_{\rm t}$, the remnant will be ejected after a single passage. For an initial orbit with $\beta < \beta_{\rm t}$, the stars can experience repeating pTDEs. The remnants in most of the cases will be fully disrupted, but in certain parameter space the remnants will be ejected after several passages.}
\label{fig:num_pTDE}
\end{figure}

By interpolating the $\rho_{\rm c}$ - $\beta$ relation in Figure \ref{fig:remnant_property}, we can calculate the evolution of $\rho_{\rm c}$ and $\beta$, and determine the number of passages for a star in an initial orbit with $\beta$ and $e$, as shown in Figure \ref{fig:num_pTDE}. Generally, a star undergoes only several passages before full disruption. More centrally concentrated stars experience more passages before full disruption. In the $\beta$ range considered here ($\beta \gtrsim 0.5$ and $0.6$ for $\gamma = 5/3$ and $4/3$, respectively), we find the total number of passages a star can undergo is generally $3$--$4$, though this number could be higher for smaller $\beta$.

Initially highly eccentric orbits ($e \sim 1$) with $\beta \gtrsim \beta_{\rm t}$ could result in ejection, transforming the stars into high-velocity stars after the passage. Repeating pTDEs can occur only for cases with $\beta \lesssim \beta_{\rm t}$ and initially relatively bound orbits. In most of the repeating pTDEs, the remnants will be fully disrupted at the end, but some of them maybe ejected after several passages.

The critical initial eccentricity separating repeating and non-repeating pTDEs is given by $\epsilon_{\rm ml} \simeq G M_{\rm h}(1-e)/(2R_{\rm p})$. From this, we find that the critical eccentricity scales as $(1-e) \propto M_{\rm h}^{-2/3} M_*^{-1/3}$. Therefore, around IMBHs the orbits needs be more bound for repeating pTDEs to happen.

It is important to note that we have not fully considered the physics involved in the long-term stellar evolution post each passage, such as heating (via nuclear reactions) and cooling (via radiation and convection), as well as viscosity. Additional heating mechanism might further pump up the remnant, but cooling mechanisms could counteract this by releasing internal energy and preventing expansion. Furthermore, if the remnant can remain tidally excited upon returning to the pericenter, and subsequent encounters can inject new excited modes into the remnant. However, these new modes might not perturb the remnant as intensely. Consequently, the central region of the remnant may remain largely intact during subsequent encounters, potentially allowing for a greater number of passages. Additionally, the stellar spin, which we do not consider here, could also affect the outcome of the following passages. Nonetheless, understanding how realistic remnants evolve through multiple passages requires further exploration in future studies.

Recent work by \cite{liu_repeating_2024} simulated multiple pTDE passages and found that stars become increasingly more vulnerable than that predicted by the adiabatic mass loss approximation after each passage. As a result, stars may only undergo several to ten passages before experiencing the full disruption. Similarly, \cite{Kiroglu_partial_2023} also found comparable results, albeit focusing on pTDEs involving much smaller IMBHs.
%%%%%%%%%%%%%%%%%%%%%%%%%%%%%%%%%%%%%%%%

%%%%%%%%%%%%%%%%%%%%%%%%%%%%%%%%%%%%%%%%

%%%%%%%%%%%%%%%%%%%%%%%%%%%%%%%%%%%%%%%%%

\section{Discussion}
\label{sec:discussion}
\subsection{Observed repeating pTDE candidates}

\begin{table*}
\centering \footnotesize
\caption{Repeating pTDE candidates.}
\label{tab:pTDE_candidates}
\begin{tabular}{ccccc}
\toprule
Candidate & Period (recurrence time) & BH mass & Num. of flares &  \\
 & (days) & ($M_{\odot}$) & & \\
\midrule
ASASSN-14ko & 114 & $7\times 10^7$ & $\gtrsim 20$ & \cite{Payne_14ko_2021,Payne_chandra_2023} \\
AT 2018fyk & 1200 & $5\times 10^7$ & 2 & \cite{Wevers_2018fyk_2023} \\
AT 2020vdq & 1000 & $4\times 10^5$ & 2 & \cite{Somalwar_first_2023} \\
eRASSt J045650.3-203750 & 223 & $10^7$ & 5 & \cite{Liu_J045650_2023,liu_rapid_2024} \\
RX J133157.6-324319.7 & 9125 & $10^6$ & 2 & \cite{Malyali_rebrightening_2023} \\
AT 2022dbl & 709 & $10^6$ & 2 & \cite{lin_unluckiest_2024}\\
\bottomrule
\end{tabular}
\tablecomments{\footnotesize In columns 3, we list their central BH mass of the host galaxy, which are obtained by the $M_{\rm h}$-$\sigma$ or bulge mass relation. Columns 4 shows their total numbers of flares that have been observed so far.}
\end{table*}

Several repeating pTDE candidates have been observed so far, as listed in Table \ref{tab:pTDE_candidates}. Their initial flares have been confirmed as TDEs through their photometric and spectroscopic properties. Subsequently, their re-brightening episodes exhibit similar observational features. Notably, ASASSN-14ko displays obvious periodic behavior, with $\gtrsim 20$ flares observed so far and ongoing activity.

The formation mechanism of these systems remains a debated question. Their periods, ranging from $100$ to $10000$ days, indicate mild eccentric orbits with $e\simeq 0.99$ to $0.999$. If they approach the SMBH on a parabolic orbit due to two-body relaxation, our calculations show that the energy deposition solely to oscillations should not be enough to form such bound systems. One possibility is tidal separation, where the stars are initially in the binary system, become tidally separated upon approaching the SMBH, with one star being bound to the SMBH while the companion is ejected \citep{Hills_Hyper_1988}.

Another question is how much flares these pTDEs can produce before full disruption. Except for ASASSN-14ko, which has exhibited over 20 flares, the other candidates have only shown 2 - 5 flares so far. They may re-brighten in the future if their remnants remain unfully disrupted. However, our simulations suggest that the remnants become increasingly vulnerable to complete disruption after tidal interaction with the BH, as tidal forces inject energy into the star, deforming and stretching it. Figure \ref{fig:num_pTDE} shows that
the total number of pTDEs is limited before complete disruption. Therefore, we do not anticipate that the repeating pTDE candidates can continue to produce more flares in the future.
However, ASASSN-14ko is an exception with flare numbers far exceeding our predictions. Furthermore, its luminosity remains relatively stable between each flare. This suggests that ASASSN-14ko may be a pTDE involving an evolved star with a compact core, allowing it to graze only the thin layer of the envelope during each passage and keeping the core largely intact \citep{MacLeod_Spoon_2013,Liu_Tidal_2023,liu_repeating_2024}.

\begin{figure}		
\centering
 \includegraphics[scale=0.5]{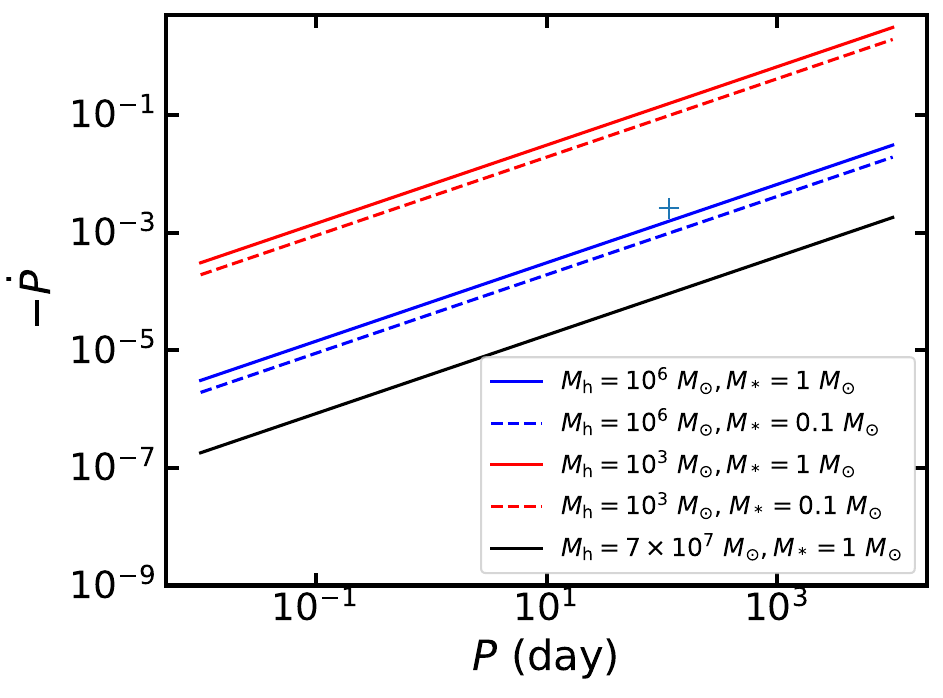}
\caption{Derivative of the remnant orbital period versus the period, based on Eq. (\ref{eq:dP_P}). This relationship assumes the orbital shrinkage is solely due to energy deposition from tidal excitation ($\epsilon_{\rm osc}$). The cross point represents the observational data for ASASSN-14ko.}
\label{fig:dP_P}
\end{figure}

It's worth noting that ASASSN-14ko exhibits a negative periodic derivative $\dot P \simeq -0.0026$, suggesting a gradual decay of its orbit. This orbital shrinkage is suspected to result from energy deposition through tidal excitation \cite{Payne_14ko_2021,Ryu_tidal3_2020}. Here we compare our prediction of energy deposition through tidal excitation to the value of this source.

The peak energy deposition as shown in Figure \ref{fig:Eorb_beta} is $\sim 10^{14}\ {\rm erg/g}$, that corresponds to a periodic derivative of
\begin{equation} \label{eq:dP_P}
    \begin{split}
    \dot P &= -\frac{3}{2} \frac{\Delta \epsilon_{\rm orb}}{\epsilon_{\rm orb}} \\
    &\sim -10^{-3} \left(\frac{P}{100\ {\rm day}}\right)^{2/3} \left(\frac{M_{\rm h}}{10^6\ {\odot}}\right)^{-2/3} \left(\frac{M_*/M_{\odot}}{R_*/R_{\odot}}\right).
    \end{split}
\end{equation}
We show it in Figure \ref{fig:dP_P} alongside the observational value of $\dot P$ in ASASSN-14ko. The observed trend aligns with expectation from tidal encounter with a $10^6\ M_{\odot}$ SMBH, suggesting that orbital shrinkage may indeed be attributed to tidal excitation. However, it is crucial to note that this result is sensitive to the central BH mass. If the BH mass is approximately $\sim 7 \times 10^7\ M_{\odot}$, as derived from properties of the host galaxy \citep{Payne_14ko_2021}, it would lead to an underestimation of the periodic derivative. It is noteworthy that other effects, e.g., star-disk interaction, can also cause the orbital shrinkage and similar periodic derivative \citep{linial_period_2024}.

\subsection{Regimes of repeating and non-repeating pTDEs}

Only a few repeating pTDE candidates have been observed so far, likely originating from the tidal separation of binary systems. In contrast, most TDE candidates, which exhibit only one flare, are believed to result from stellar two-body relaxation. The search for X-ray and optical flares from tidal disruption events has identified over 100 TDE candidates \citep{gezari_review_2021}. Some researches have focused on the so-called ``rate problem", where the observed rate of TDE detections in most galaxies, $\sim 10^{-5}\ {\rm yr^{-1}}$ \citep{donley_large-amplitude_2002,velzen_measurement_2014}, seems roughly a few to an order of magnitude lower than the prediction of TDE rate based on loss-cone theory \citep{wang_revised_2004,Stone_Rates_2016,pfister_enhancement_2020,roth_forward_2021,Bortolas_pTDE_2023,chang_rates_2024}.

Most studies consider the loss-cone regime, within which a star will plunge into the tidal radius $R_{\rm t}$ (or up to $\sim 2 R_{\rm t}$ for pTDE) and produce a TDE flare. Full TDEs typically occur in the full loss-cone regime, where the two-body relaxation timescale is shorter than the orbital period, allowing the star's trajectory to deeply penetrate the tidal radius. Conversely, in the empty loss-cone regime, most stars only graze the marginal mass-loss radius of $\sim 2 R_{\rm t}$, producing week pTDEs. These weak pTDEs might be too dim to detect, potentially alleviating the ``rate problem" tension. \citep{Bortolas_pTDE_2023,broggi_repeating_2024}.

Even if these weak pTDEs are not detectable, the remnant could return with energy deposition after the encounter and be tidally disrupted again. As shown in Section \ref{sec:pTDE}, the remnant becomes more vulnerable to disruption, leading subsequent encounters to strip more mass and produce more luminous flares. After several encounters, the star may be fully disrupted or ejected if $\beta$ is slightly larger.
 
For a star in a marginally mass-loss orbit ($\beta \sim 0.5$), if the orbit-averaged perturbation is minimal, the star could continue to pass through the pericenter until it is fully disrupted. However, if the perturbation is significant enough to shift the pericenter past the transition point $\beta > \beta_{\rm t}$, the remnant will be ejected or fully disrupted. 

Usually, the filling factor $q \simeq (\delta j/j_{\rm lc})^2$ is used to characterize the loss-cone regime, where $\delta j$ is the orbital angular momentum change in one orbit due to two-body perturbations, and $j_{\rm lc} = \sqrt{2GM_{\rm h}R_{\rm t}}$. $q$ is a function of the separation from the central MBH (see Figure 3 of \cite{Bortolas_pTDE_2023} for the Milky way model). Stars close to the central MBH are in the empty loss-cone ($q \ll 1$) regime, where they are unlikely to experience large orbital changes due to small perturbation. Therefore, if their pericenters are close enough for pTDEs, they are likely to return and undergo multiple passages, making repeating pTDEs more probable in the empty loss-cone regime.

\begin{figure}		
\centering
 \includegraphics[scale=0.4]{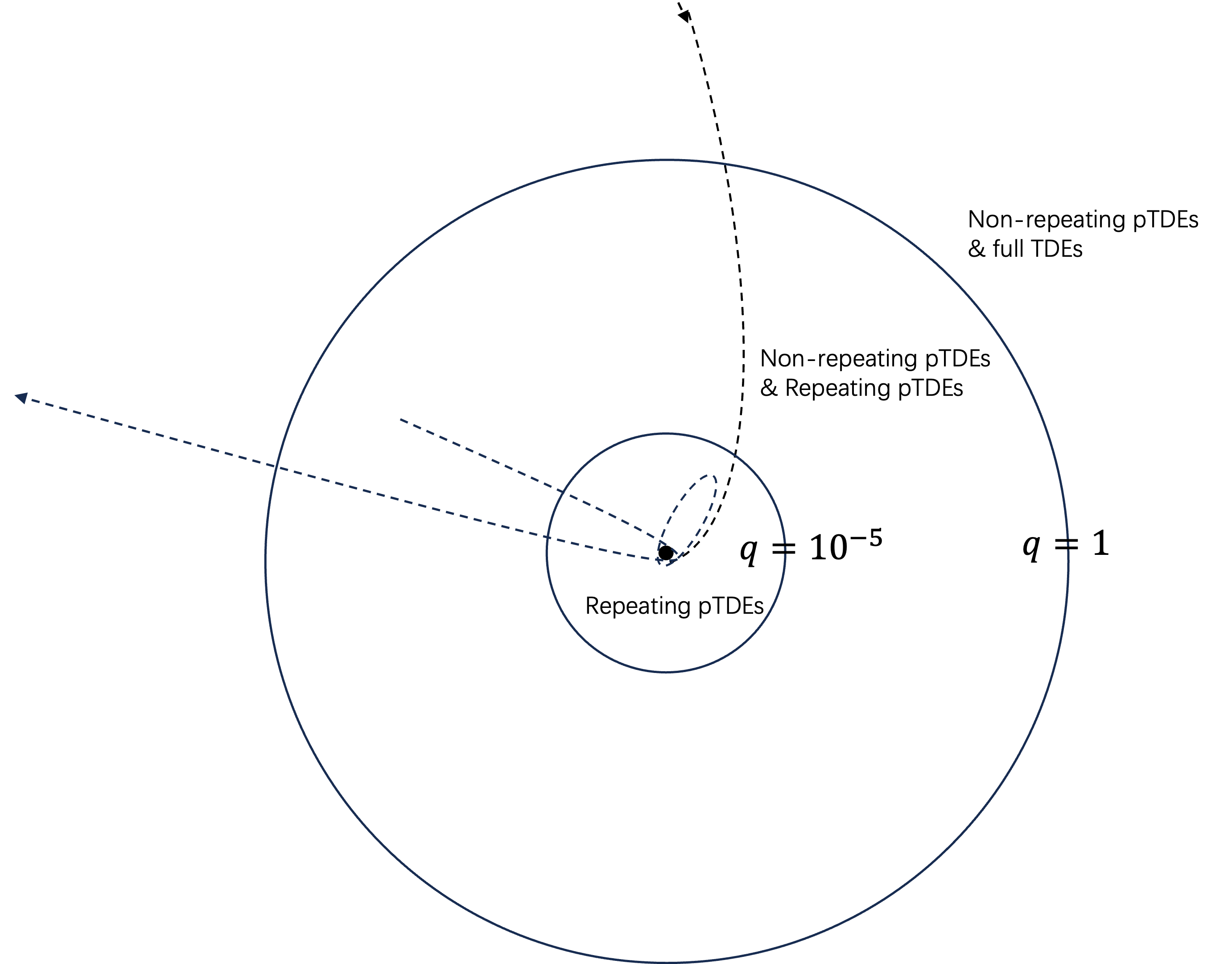}
\caption{Sketch of the loss-cone dynamics for a $10^6\ M_{\odot}$ SMBH system. The diagram includes two circles that delineate different regimes: the inner circle represents the extreme empty loss-cone regime with $q\lesssim 10^{-5}$, while the outer circle ($q=1$) marks the boundary between the full and empty loss-cone regimes. Repeating pTDEs are likely to occur safely within the extreme empty loss-cone regime. In the intermediate regime ($10^{-5} \lesssim q \lesssim 1$), most stars initially have small $\beta$. However, after one or two weak pTDEs, their $\beta$ would increase beyond $\beta_{\rm t}$, leading to their ejection from the system, as shown in Figure \ref{fig:num_pTDE}. For the smaller fraction of stars with larger initial $\beta$, the first pTDE with $\beta \gtrsim \beta_{\rm t}$ would eject the remnant, resulting in some non-repeating events. In the full loss-cone regime, primarily full TDEs and non-repeating pTDEs are expected to occur.}
\label{fig:loss_cone}
\end{figure}

However, stars in the empty loss-cone regime with $\beta > \beta_{\rm t}$ would be ejected after a single passage. As shown in Figure \ref{fig:num_pTDE}, to safely assume the occurrence of repeating pTDEs, preventing ejection in the first few passages, the orbit should be less eccentric with $(1-e) \gtrsim 10^{-4}$, corresponding to a pericentre $R_{\rm p} \gtrsim 10^{-4} a$, which is smaller than the tidal disruption radius when $a \lesssim 0.02$ pc for a $10^6\ M_{\odot}$ SMBH. In Figure 3 of \cite{Bortolas_pTDE_2023}, this orbital separation corresponds to an extreme empty loss-cone regime with $q\lesssim 10^{-5}$. In this regime, the relaxation is also negligible for the pericenter change, because $q \simeq \delta R_{\rm p}/R_{\rm p} \simeq -\delta \beta / \beta \lesssim 10^{-5}$. We illustrate this in Figure \ref{fig:loss_cone} to show in which regime the stars can undergo repeating/non-repeating pTDEs and full TDEs.

Therefore, only some of the pTDEs in the empty loss-cone regime can be repeating. Some repeating pTDEs with $\beta \lesssim \beta_{\rm t}$ are weak and might be too dim to detect. The orbital change in pTDEs should significantly affect the overall rate of TDEs.

\subsection{Emission properties of pTDEs}
TDE candidates observed so far can be categorized into two main types: optical and X-ray bright TDEs. This distinction also applies to the few known repeating pTDE candidates as shown in Table \ref{tab:pTDE_candidates}). For instance, ASASSN-14ko, AT 2020vdq and AT2022dbl are optical bright repeating pTDEs, while other candidates are identified by their repeating bright X-ray emissions. 

The origin of emissions in TDEs has been a subject of long-standing debate. It is widely accepted that the X-ray emission, with temperature of $\sim 10^5$ K, originates from the inner part of the accretion disk. However, the optical/UV emission, which has a lower temperature of $\sim 10^4$ K and evolves differently from the X-ray emission, remains more elusive. Several theories suggest that the optical/UV emission may stem from the circularization process driven by stream-stream (or stream-disk) collisions  \citep{Piran_DISK_2015,Chen_pTDE_2021,steinberg_stream-disk_2024}, arise from a reprocessing layer produced during circularization process \citep{Jiang_PROMPT_2016,Lu_Self_2020}, or be driven by the super-Eddington accretion process within the accretion disk \citep{Strubbe_Optical_2009,Lodato_Multiband_2011,Metzger_A_2016,dai_unified_2018,Thomsen_Dynamical_2022}.

In the following discussion, we will explore how changes in the remnant’s orbit affect the emission properties of pTDEs. Two primary mechanisms for emission are considered: disk accretion power and stream-stream collision. If the accretion disk efficiently accretes the fallback mass, the accretion would follow the mass fallback rate, with the bolometric luminosity given by $L_{\rm acc} \simeq 0.1 \dot{M}_{\rm fb} c^2$, assuming an efficiency of $0.1$. If the emission is produce by the stream-stream collision, the peak luminosity can be estimated using the fomula provided by \citep{Chen_pTDE_2021}:
\begin{equation} \label{eq:Lcir}
    \begin{split}
    L_{\rm cir} &\simeq 6 \times 10^{42}\ \beta^{9/2} \left(\frac{\Delta M}{0.01\ M_{\odot}}\right) \left(\frac{M_{\rm h}}{10^6\ M_{\odot}}\right)^2 \\ 
    &\times \left(\frac{M_*}{M_{\odot}}\right)^{3/2} \left(\frac{R_*}{R_{\odot}}\right)^{-9/2}\ {\rm erg\ s^{-1}}.
    \end{split}
\end{equation}

\begin{figure}		
\centering
 \includegraphics[scale=0.5]{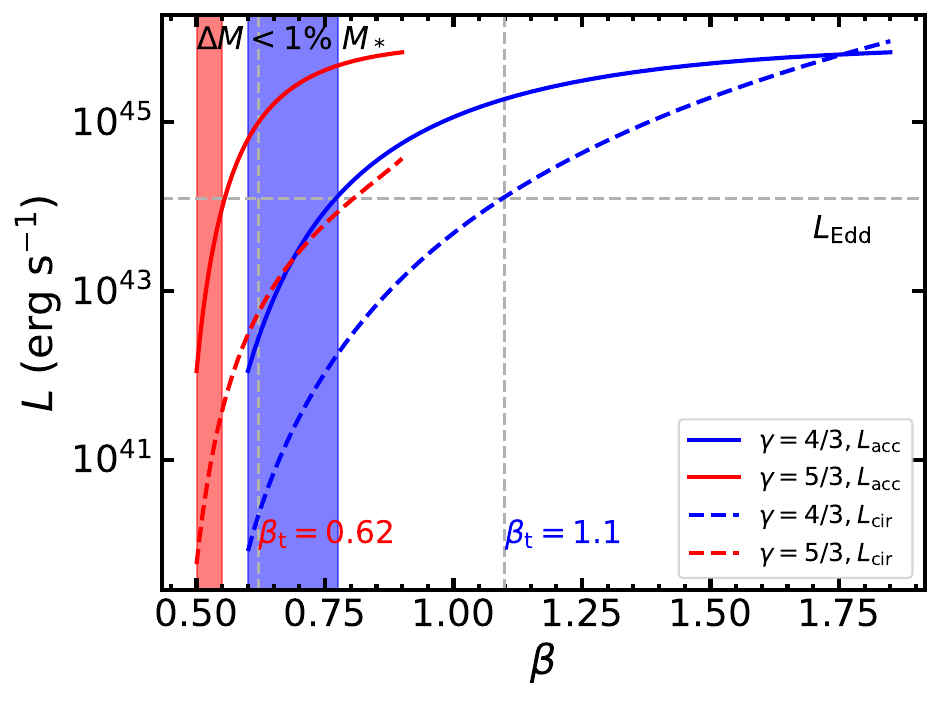}
\caption{Expected luminosity of pTDE as a function of $\beta$ ($M_*=M_{\odot}$, $R_*=R_{\odot}$ and $M_{\rm h}=10^6\ M_{\odot}$). The solid and dashed lines represent the peak luminosities from accretion power $L_{\rm acc} \simeq 0.1 \dot{M}_{\rm peak} c^2$ and the stream circularization $L_{\rm cir}$ as given by Eq. (\ref{eq:Lcir}), respectively. The horizontal dashed gray line marks the Eddington luminosity, while the vertical dashed gray lines represent $\beta_{\rm t}$. The colored shaded regions denote where the stripped mass is less than $1\%\ M_*$.}
\label{fig:emission}
\end{figure}

Using the fitting formulas from \cite{Guillochon_Hydrodynamical_2013}, we calculate $\Delta M$ and peak mass fallback rate $\dot{M}_{\rm peak}$, and subsequently determine $L_{\rm acc}$ and $L_{\rm cir}$ at peak. The results are shown in Figure \ref{fig:emission}.

We find that the accretion luminosity becomes super-Eddington when the stripped mass exceeds $\gtrsim 0.01\ M_*$. In such cases, pTDEs could produce disk winds and reprocess X-ray photons into optical/UV emission, similar to the scenario observed in full TDEs. Consequently, the optical light curves and spectra of pTDEs would resemble those of full TDEs, which is consistent with the observational characteristics of ASASSN-14ko, AT 2020vdq, and AT 2022dbl. However, the light curve’s decay could be steeper than the typical  $t^{-5/3}$ power-law decay seen in full TDEs \citep{Guillochon_Hydrodynamical_2013,coughlin_partial_2019,Ryu_tidal3_2020}. For stars in eccentric orbits, the light curve’s decay might exhibit a steep drop rather than a gradual power-law decline \citep{hayasaki_finite_2013}. In the case of AT 2020vdg \citep{Somalwar_first_2023}, The second flare has a steeper decay than the first one, which might be attributed to an decease in the orbit's eccentricity during the second encounter.

For very small $\beta$ where $\Delta M \lesssim 0.01\ M_*$, the accretion would be sub-Eddington, resulting in primarily low-luminosity X-ray emission. If these weak pTDEs are repeating, they could contribute to the variability of the X-ray background emission, such as that observed in low-luminosity AGNs. Sub-Eddington repeating pTDEs with $\beta < \beta_{\rm t}$ will experience orbital shrinkage until they eventually undergo a full TDE.  

%%%%%%%%%%%%%%%%%%%%%%%%%%%%%%%%%%%
\section{Conclusion}
\label{sec:conclusion}
In this paper, we conduct dozens simulations  using the FLASH code to thoroughly explore the pTDE of a star by a SMBH or an IMBH, ranging from minimal mass loss to full disruption. Our primary focus is on the remnant's orbit and structure after the first encounter. Our key findings are:
\begin{enumerate}
    \item We confirm the underlying physical mechanism for the orbital change of the remnant: energy deposition into tidal oscillatory modes causes the remnant to lose orbital energy, while asymmetric mass loss causes the remnant to gain energy.
    
    \item We determine the $\beta$ ranges for the remnant to lose or gain orbital energy. We find that for $\beta \lesssim 0.62$ and $\beta \lesssim 1.1$ for $\gamma = 5/3$ and $\gamma = 4/3$ polytropic stars, respectively, the remnant will lose orbital energy. Conversely, for higher $\beta$ values, the remnant will gain energy. This orbital change dictates the remnant's fate -- whether it returns to the pericenter to be disrupted again or is ejected.
    
    \item The orbital energy change of the remnant, expressed in Eq. (\ref{eq:dE_orb2}), depends only on the stellar structure and $\beta$. In the limit of $M_{\rm h} \gg M_*$, $\Delta \epsilon_{\rm orb}$ is independent of $M_{\rm h}$. These findings are applicable to various pTDEs involving different stars and MBHs.

    \item For a star initially in a parabolic orbit, the remnant would return to the pericenter and be disrupted again in approximately 30 years for a $10^3\ M_{\odot}$ IMBH, unless the remnant orbit is perturbed by interactions with other stars during the second orbit. However, for a SMBH, the return time is very long, around 10,000 years, making it unlikely to detect recurrent pTDEs in such systems within the lifetime of current surveys.
    
    \item We also study the structure of the remnant. It consists of a central region that rotates almost rigidly and a thin, extended envelope that rotates differentially near the break-up velocity. This structure differs significantly from that of a typical star. We can search for these kinds of remnants in galactic nuclei, and the morphology of six G-objects detected in our galactic center is similar to that of these remnants.
    
    \item Using the properties of the remnant and the dependence of orbital changes, we predict subsequent passages until final full disruption or ejection. After each encounter, the injected energy makes the remnant more vulnerable, resulting in a total number of passages fewer than ten. A more centrally concentrated star will have more passages before full disruption.
\end{enumerate}

These results are crucial for understanding the pTDEs, predicting their evolution and observational features, as well as more accurate estimates of event rates.

Many upcoming or newly operational time-domain telescopes will perform sensitive all-sky surveys, likely detecting more TDEs, especially recurrent events, and providing insights into the underlying physics of TDEs. The Rubin Observatory will offer a deep and sensitive optical survey, potentially discovering more weak pTDEs. The X-ray survey telescope Einstein Probe can detect X-ray flares from TDEs. By comparing with archival data, we can find more repeating pTDE candidates over long periods. Additionally, the GRAVITY, SINFONI, and Keck observatories have the potential to identify more remnant candidates from pTDEs in the galactic center.

%%%%%%%%%%%%%%%%%%%%%%%%%%%%%%%%%%%
\begin{acknowledgments}
We thank Enrico Ramirez-Ruiz for the useful discussion. J.H.C thanks Xinghao Chen for the discussion of stellar oscillation. J.H.C. and L.D. acknowledge the support from the National Natural Science Foundation
of China and the Hong Kong Research Grants Council (HKU12122309, 17314822, 17304821). S.-F. Liu acknowledges the support from the Guangdong Basic and Applied Basic Research Foundation under grant No. 2021B1515020090, the National Natural Science Foundation of China under grant No. 11903089, and the China Manned Space Project under grant Nos. CMS-CSST-2021-A11 and CMS-CSST-2021-B09. J.-W. Ou acknowledges the support from the Talent Introduction Program of Shaoguan University (440-9900064601), and the Key Project of the Natural Science Research of Shaoguan University (SZ2021KJ10). The FLASH code used for the simulations in this work was developed in part by the DOE NNSA and the DOE Office of Science-supported Flash Center for Computational Science at the University of Chicago and the University of Rochester. The authors would also like to thank the National Supercomputing Center in Guangzhou Center and Nansha Subcenter for providing high-performance computational resources.
\end{acknowledgments}

%%%%%%%%%%%%%%%%%%%%%%%%%%%%%%%%%%%%%%%%%

%%%%%%%%%%%%%%%%%%%%%%%%%%%%%%%%%%%%%%%%%%%%%%%%%%%%%
\bibliography{cited}

\end{CJK*}
\end{document}